\documentclass[%
reprint,
showpacs,
showkeys,
amsmath,amssymb,aps,pra,
]{revtex4-1}
\usepackage{graphics}% Include figure files
\usepackage{epsfig}
\usepackage{dcolumn}% Align table columns on decimal point
\usepackage{color}
\usepackage{bm}% bold math
\usepackage{soul}% crossing out
\begin{document}

%\preprint{APS/123-QED}

\title{Electric manipulation of the Mn-acceptor binding energy 
and the Mn-Mn exchange interaction on
the GaAs (110) surface by nearby As vacancies}% Force line breaks with \\

\author{M.~R.~Mahani$^{1}$,  A.~H.~MacDonald$^{2}$, and C.~M.~Canali$^{1}$}

\affiliation{$^{1}$Department of Physics and Electrical engineering, Linnaeus University, 391 82 Kalmar, Sweden}
\affiliation{$^{2}$Department of Physics, University of Texas at Austin, Austin, Texas 78712, USA}

\date{\today}% It is always \today, today,

\begin{abstract}
We investigate theoretically the effect of nearby As (arsenic) vacancies on the magnetic properties of 
substitutional Mn (manganese) impurities on the GaAs (110) surface, using a 
microscopic tight-binding model which captures the salient features of the
electronic structure of both types of defects in GaAs. 
The calculations show that the binding 
energy of the Mn-acceptor is essentially unaffected by the presence 
of a neutral As vacancy, even at the shortest possible ${\rm V}_{\rm As}$--Mn separation.
On the other hand, in contrast to a simple tip-induced-band-bending theory and in 
agreement with experiment, for a positively charged 
As vacancy the Mn-acceptor binding energy is significantly reduced as the As vacancy is brought closer to the Mn impurity.
For two Mn impurities aligned ferromagnetically, we find that 
nearby charged As vacancies enhance the energy level splitting of the associated coupled acceptor levels, 
leading to an increase of the effective exchange interaction. Neutral vacancies leave the exchange splitting
unchanged. Since it is experimentally possible to switch reversibly between the two charge states of the vacancy,
such a local electric manipulation of the magnetic dopants could result in an efficient real-time control of their exchange interaction.
\end{abstract}

\pacs{75.50.Pp,
71.55.-i %defect states
}% PACS, the Physics and Astronomy Classification Scheme.
\keywords{As vacancy, Mn-acceptor binding energy, Mn-pair exchange splitting}%Use showkeys class option if keyword

\maketitle

%\tableofcontents

\section{\label{sec:level1}INTRODUCTION\protect}

State-of-the-art scanning tunneling microscopy (STM) is nowadays
able to manipulate and probe the physical properties of individual 
impurities located near the surface of a semiconductor. 
Scanning tunneling spectroscopy (STS) along with STM  provides 
excellent spatial and energy resolution down to the atomic scale. These techniques 
have been used to study the magnetic and electronic 
properties of individual Mn impurities in GaAs~\cite{yakunin_prl04, yakunin_prl05},
and visualize the topography of the acceptor state bound to Mn on
the GaAs (110) surface~\cite{yazdani_nat06,yazdani_JAP_07}. 
A more recent advance of these techniques 
employs the electric field generated by an As vacancy in GaAs to influence
the environment surrounding substitutional Mn impurities in the host 
material~\cite{gupta_science_2010}. By using STM, it is possible to 
place As vacancies in GaAs with atomic precision, switch the 
vacancy between its charge states, and address the effect of
the As vacancy electrostatic field on individual Mn-acceptors.\\
The electronic structure and charge state of As vacancies
in GaAs have been investigated theoretically and experimentally 
over the past two decades~\cite{zhang_PRL_1996, 
northrup_prb_94, zhang_prl91, chelikowsky_prl96, 
seong_PRB_1995, daw_prb79, lengel_jvc93, aloni_jcp01, 
lengel_prl94, Cheli_surf.sci_98, yi_prb95, chao_prb96}.
The most stable charge state of As vacancies has been the topic of a long debate. 
For dopants in bulk GaAs, the {\it ab-initio} work in Refs.~\onlinecite{zhang_prl91, northrup_prb_94} 
showed that the $(+3)$ charge state in $p$-type GaAs
is the most stable charge state of an As vacancy, while this value is 
$(+1)$ for $n$-type GaAs. This result was confirmed later by tight-binding calculations~\cite{seong_PRB_1995}.
Determining the stable charge state of an As vacancy on the (110) 
surface of GaAs turned out to be more challenging and, for a while, contradictory conclusions were reached.  
STM experiments along with tight-binding calculations~\cite{lengel_prl94} show that $(+2)$
is the stable charge state of an As vacancy on the $p$-type 
GaAs (110) surface, while a pseudopotential
calculation~\cite{yi_prb95} found a charge state  $(-1)$ instead. 
However, a more careful {\it ab-initio} study of
As vacancies on the GaAs (110) surface~\cite{zhang_PRL_1996} 
suggested $(+1)$ as the stable charge state.  Further support for the $(+1)$ charge state as
the most stable state
of As vacancies on the $p-$type GaAs (110) 
surface was provided later by {\it ab-initio} work~\cite{chao_prb96, chelikowsky_prl96, Cheli_surf.sci_98} and
by recent STM experiments~\cite{gupta_science_2010}. 
Based on this evidence, among the possible charged states of the As vacancies in GaAs,
in this paper we will consider the $(+1)$ state, together with the neutral state.\\ 
In addition to the charge state, the electronic structure of As vacancies
has also been investigated~\cite{chelikowsky_prl96, Cheli_surf.sci_98}.
Kim and Chelikowsky~\cite{chelikowsky_prl96} studied  the
electronic structure of As vacancies on the GaAs (110) surface using {\it ab-initio} methods.
Their calculations show the presence of three distinct doubly-degenerate vacancy levels,
one of which is fully occupied and lies in the
valence band, while the other two levels are within the GaAs band gap. The occupancy of these two levels depends on the charge state of 
the As vacancy: for a neutral vacancy the lowest of the two is half-filled, while the other is empty. For a (+1) charge state, both levels
are empty.
The double-degeneracy of these levels, 
which is a trivial spin-degeneracy if spin-orbit interaction is neglected,
is preserved in the presence of spin-orbit as a Kramer's degeneracy 
as a consequence of time-reversal symmetry.
In a second publication~\cite{Cheli_surf.sci_98} the same authors provided a more thorough analysis of the  
single-particle energy levels of substitutional As vacancies placed on  a $p-$type GaAs (110) surface, 
together with theoretical STM topographies of As vacancies.\\
Before reviewing the experimental work on the effect of 
As vacancies on the Mn-acceptor states
in GaAs~\cite{gupta_science_2010}, we would like to 
mention relevant studies on Mn near the GaAs (110) surface.
Mn dopants in GaAs have been extensively studied theoretically by first-principles 
calculations~\cite{zhao_apl04,sarma_prl04,
PhysRevB.70.085411, fi_cmc_prb_2012}, 
microscopic tight-binding (TB) 
models~\cite{tangflatte_prl04, tangflatte_prb05, timmacd_prb05,Jancu_PRL_08, 
scm_MnGaAs_paper1_prb09, 
scm_MnGaAs_paper2_prb2010, scm_MnGaAs_paper3_prl011, 
PhysRevLett.105.227202, mc_MF_2013}, 
as well as  experimentally by STM techniques~\cite{yakunin_prl04, 
yakunin_prl05, yazdani_nat06, yazdani_prb09, PhysRevB.75.195335}. 
Among theoretical studies, TB models are computationally efficient, which 
is valuable for studies of impurity problems because of the necessity of large supercell sizes.  
They have been particularly successful in describing the complex electronic and magnetic properties of some TM 
impurities, such as Mn dopants, and their associated acceptor states~\cite{tangflatte_prl04, tangflatte_prb05, 
Jancu_PRL_08, scm_MnGaAs_paper1_prb09, scm_MnGaAs_paper3_prl011, mc_MF_2013, PhysRevB.89.165408}.
In the case of a pair of Mn, 
TB calculations~\cite{yazdani_nat06, tang_spie2009, scm_MnGaAs_paper2_prb2010} 
have provided insight
into the magnetic interactions between substitutional Mn ions near the surface of GaAs.
The high-resolution STM measurements in Refs.~\cite{yazdani_nat06,yazdani_JAP_07}, 
studied a pair of ferromagnetically coupled Mn atoms on the GaAs surface, 
demonstrating~\cite{yazdani_nat06, tang_spie2009, scm_MnGaAs_paper2_prb2010} 
that the splitting between the two 
acceptor levels measured by STM is related to the exchange interaction between the Mn atoms.\\
Control over the acceptor binding energy of a substitutional Mn dopant in GaAs by charged vacancies 
was recently studied in Ref.~\onlinecite{gupta_science_2010}.
The authors were able to use a STM tip to place As vacancies in the vicinity of Mn dopants,
and switch reversibly the charge state of the As vacancy (from a $(+1)$ state to a neutral state) by applying
a small voltage pulse.
Although a neutral vacancy did not change the binding energy of
the Mn-acceptor, a positively charged vacancy was found to
decrease the Mn-acceptor binding energy by an amount that increases with decreasing
As-vacancy--Mn-dopant separation.\\
In this paper we provide a theoretical simulation of neutral and charged As vacancies,
providing the electronic properties of As vacancies in agreement with previous studies.
The effects of As vacancies on the Mn-acceptor binding energy
have been calculated theoretically, and the results are compared with recent experiment.
In addition to one Mn impurity, our calculations predict the enhancement of the acceptor-splitting between a pair
of ferromagnetically coupled Mn on a GaAs (110) surface.\\
The paper is divided into the following sections. In Sec.~\ref{theo_model}, we introduce 
a theoretical approach to model substitutional Mn impurities 
in the presence of neutral and charged As vacancies on a GaAs (110) surface, extending the TB model
for Mn impurities in GaAs originally introduced in Ref.~\onlinecite{scm_MnGaAs_paper1_prb09}.
In Sec.~\ref{As_vac}, we present results for the electronic properties
of an As vacancy on a GaAs (110) surface. 
Among all the possible charged states of As vacancies 
on GaAs (110) surfaces,
we consider only the $+1$ state, ${\rm V}_{\rm As}^+$, 
which is the most stable in $p$-doped samples,  
together with the neutral state, ${\rm V}_{\rm As}^0$. 
In the Sec.~\ref{As_vac_Mn} we use the theoretical model to investigate
the effect of ${\rm V}_{\rm As}^0$ and
${\rm V}_{\rm As}^+$ on the properties of
an individual substitutional Mn. We focus on the physical mechanism leading
to the reduction of the Mn-acceptor binding energy caused by nearby charged As vacancies.
In Sec.~\ref{As_vac_2Mn} we investigate the effect of As vacancies
on the exchange interaction between two ferromagnetically 
coupled Mn dopants on the GaAs (110) surface. Based on the results of our calculations,
we predict that the energy splitting between two coupled Mn-acceptor states should increase 
under the electrostatic potential of a nearby positively charged vacancy.  This prediction should 
be experimentally verifiable. Our calculations also reveal that nearby As vacancies introduce
unoccupied mid-gap impurity levels, in addition to those the ones associated with the Mn-acceptor only.
These extra states are coupled in a non-trivial way with the Mn states, and should also
be visible as additional peaks in STM measurements.
Finally in Sec.~\ref{conc.}, we present our conclusions and we discuss the possibility
of using vacancy-induced local electric fields to manipulate the quantum exchange interaction
between pairs of magnetic dopants.\\
\section{THEORETICAL MODEL}
\label{theo_model}
In this section we review the theoretical tight-binding (TB) model 
introduced in Ref.~\onlinecite{scm_MnGaAs_paper1_prb09}
to describe substitutional Mn impurities in GaAs,
and complement it with an additional new term modeling neutral (${\rm V}_{\rm As}^0$) or positively charged 
(${\rm V}_{\rm As}^+$) As vacancies.

The total TB Hamiltonian is 
\begin{align}
H &  =H_{\rm MnGaAs}+H_{{\rm V}_{\rm As}^{0/+}}.
\label{hamiltonian1}
\end{align}
The Hamiltonian for Mn dopants in a GaAs host crystal, $H_{\rm MnGaAs}$, is 
\begin{align}
H_{\rm MnGaAs} &  =\sum_{ij,\mu\mu^{\prime},\sigma}t_{\mu\mu^{\prime}}^{ij}a_{i\mu\sigma
}^{\dag}a_{j\mu^{\prime}\sigma}+J_{pd}\sum_{m}\sum_{n[m]}\vec{S}_{n}\cdot
\hat{\Omega}_{m}\nonumber\\
&  +\sum_{i,\mu\mu^{\prime},\sigma\sigma^{\prime}}\lambda_{i}\langle\mu
,\sigma|\vec{L}\cdot\vec{S}|\mu^{\prime},\sigma^{\prime}\rangle a_{i\mu\sigma
}^{\dag}a_{i\mu^{\prime}\sigma^{\prime}}\nonumber\\
&  +\frac{e^{2}}{4\pi\varepsilon_{0}\varepsilon_{r}}\sum_{m}\sum_{i\mu\sigma
}\frac{a_{i\mu\sigma}^{\dag}a_{i\mu\sigma}}{|\vec{r}_{i}\mathbf{-}\vec{R}_{m}|}
+V_{\rm Corr}\;.
\label{hamiltonian}
\end{align}
Here $i$ and $j$ are atomic indices for Ga and As atoms, $m$ runs over
Mn atoms, and $n[m]$ is an index labeling the As atoms that are nearest neighbors of the Mn dopants. 
The orbital indices $\mu$, $\nu$
describe a $sp^3$ model; $\sigma$ is the spin index. The first term in
Eq.~(\ref{hamiltonian}) contains the near-neighbor Slater-Koster tight-binding
parameters~\cite{slaterkoster_pr54,papac_jpcm03}. For surface calculations, we rescale the parameters
to account for surface relaxation~\cite{chadi_prl78,chadi_prb79,scm_MnGaAs_paper1_prb09}.\\
The second term implements the antiferromagnetic exchange coupling between 
the Mn spin, $\hat{\Omega}_{m}$, treated as a classical vector, and the quantum spin of the nearest neighbor 
As $p-$orbitals ($\vec{S}_n =  1/2\sum_{\pi\sigma{\sigma'}} a_{n\pi\sigma}^\dagger 
\vec{\tau}_{\sigma{\sigma'}} a_{n\pi{\sigma'}}$). Our $sp^3$ TB 
Hamiltonian does not explicitly include Mn $d-$orbitals. We use instead this effective exchange interaction term
to capture the effect of the hybridization between Mn $d-$orbitals
and the nearest neighbor 
As $p-$orbitals. 
The choice of the exchange coupling constant ($J_{pd}=1.5~eV$) has been inferred from
theory~\cite{timmacd_prb05} and experiment~\cite{ohno_sci98}. The merits and limitations of this
effective model have been discussed in detail in a recent publication\cite{PhysRevB.89.165408}.\\
The next term is the spin-orbit interaction with the spin-orbit splitting parameters taken from
Ref.~\onlinecite{chadi_prb77}. The interplay between the spin-orbit coupling and exchange interaction is the 
origin of the dependence of the acceptor wavefunction and level spectra on the direction of 
the Mn magnetic moment 
($\hat{\Omega}_{m}$)\cite{tangflatte_prl04,tangflatte_prb05, scm_MnGaAs_paper1_prb09, mc_MF_2013, PhysRevB.89.165408}.\\
The fourth term is a long-range repulsive Coulomb potential which captures the electrostatic potential produced by the Mn ion.
This potential tends to repel electrons and prevents extra electrons from approaching 
the impurity atom and therefore, prevents it from charging. This long-range potential
is dielectrically screened by the host. 
The screening constant on the surface is reduced approximately by a factor of two with respect to the bulk value
$\varepsilon_{r} = 12$. 
This crude representation of the reduced screening at the surface is qualitatively supported by
experimental results~\cite{garleff_prb_2010}.\\
The last term $V_{\rm Corr}$ consists of two parts: an on-site potential influencing
the Mn ion and an off-site term one affecting its nearest neighbors. The correction term for the Mn
is estimated to be $\approx 1.0$~eV from the ionization energy of Mn. 
We choose the value of the nearest neighbor off-site potential by tuning the position of the
acceptor level to be at the experimentally observed
position~\cite{schairer_prb74,lee_ssc64,chapman_prl67,linnarsson_prb97}.
The acceptor lies at 113~meV above the valance band for Mn in the bulk 
and $\approx 850$~meV above the valence band for Mn on the surface. 
This model successfully reproduces most of the electronic and magnetic properties of 
Mn in GaAs~\cite{scm_MnGaAs_paper1_prb09}.\\
The second term in Eq.~\ref{hamiltonian1} is a one-body effective potential describing the As vacancy. 
In order to model a neutral As vacancy, ${\rm V}_{\rm As}^0$, we add a large positive on-site energy
on the As site where the vacancy resides. This 
large on-site energy gives rise to ${\rm V}_{\rm As}^0$ impurity states in the gap.
In the case of a charged vacancy, ${\rm V}_{\rm As}^+$, we introduce the additional term
\begin{align}
H_{{\rm V}^+_{\rm As}}=-\frac{e^{2}}{4\pi\varepsilon_{0}\varepsilon_{r}}\sum_{i\mu\sigma
}\frac{a_{i\mu\sigma}^{\dag}a_{i\mu\sigma}}{|\vec{r}_{i}\mathbf{-}\vec{r}_{\rm vac}|}.
\label{ham-vac}
\end{align}
Eq.~(\ref{ham-vac}) represents an attractive potential for electrons
and is only added to the Hamiltonian when the charge state of vacancy is positive. 
It is in fact a long-range potential for a positive point charge, dielectrically screened
by the host in a way similar to the procedure explained above for the Mn ion, 
which gives rise to the ${\rm V}_{\rm As}^+$ impurity states in the gap.\\
The electronic and magnetic properties of a single 
substitutional Mn atom in the presence of an As vacancy
are obtained by performing supercell type calculations on finite clusters. 
Based on our computational resources, typically we are able to  
fully diagonalize and obtain the entire eigenvalue spectrum of 
the one-particle Hamiltonian in Eq.~\ref{hamiltonian1} for clusters 
with up to 3200 atoms. 
For surface calculations we apply periodic boundary conditions in the two (110) surface plane directions.
This corresponds to a $38\times 38\; {\rm A}^2$ surface and a supercell cluster that has 20 atomic layers along the surface normal
separating the two (110) surfaces\cite{scm_MnGaAs_paper1_prb09}.\\ 
In order to reach large enough separations between the Mn impurities and the As vacancies on the surface,
and model the experimental situation, larger clusters with larger surface areas are needed. In this case
it is not feasible to fully diagonalize the Hamiltonian.
Therefore for clusters larger than 3200 atoms, 
we used instead the Lanczos method, which allowed us to compute eigenvalues 
in an energy window of interest~\cite{JPCM_scann_p_m}, typically one
centered inside the gap of the host material,
where the important impurity levels are located. 
With this method we were able to consider clusters
containing up to 8840 atoms, whose dimensions in terms of atomic layers are 20$\times$34$\times$52
along $[110]\times[1\overline{1}0]\times[001]$ crystalline directions. These clusters were
large enough to allow us to simulate As vacancy--Mn-impurity separations of the order of 4 nm.
The outputs of the two methods (full diagonalizations vs Lanczos) 
were systematically compared for clusters up to 3200 atoms,
to ensure the reliability of the Lanczos diagonalization procedure.
We relaxed the position of atoms on the (110) surface layer following a procedure introduced in 
Refs.~[\onlinecite{chadi_prl78, chadi_prb79}] to remove artificial dangling-bond states.\\
\section{RESULTS AND DISCUSSION}
\label{the_results}
\begin{figure*}[htp]
\begin{minipage}[h]{1.0\linewidth}
\centering
\includegraphics[scale=0.32]{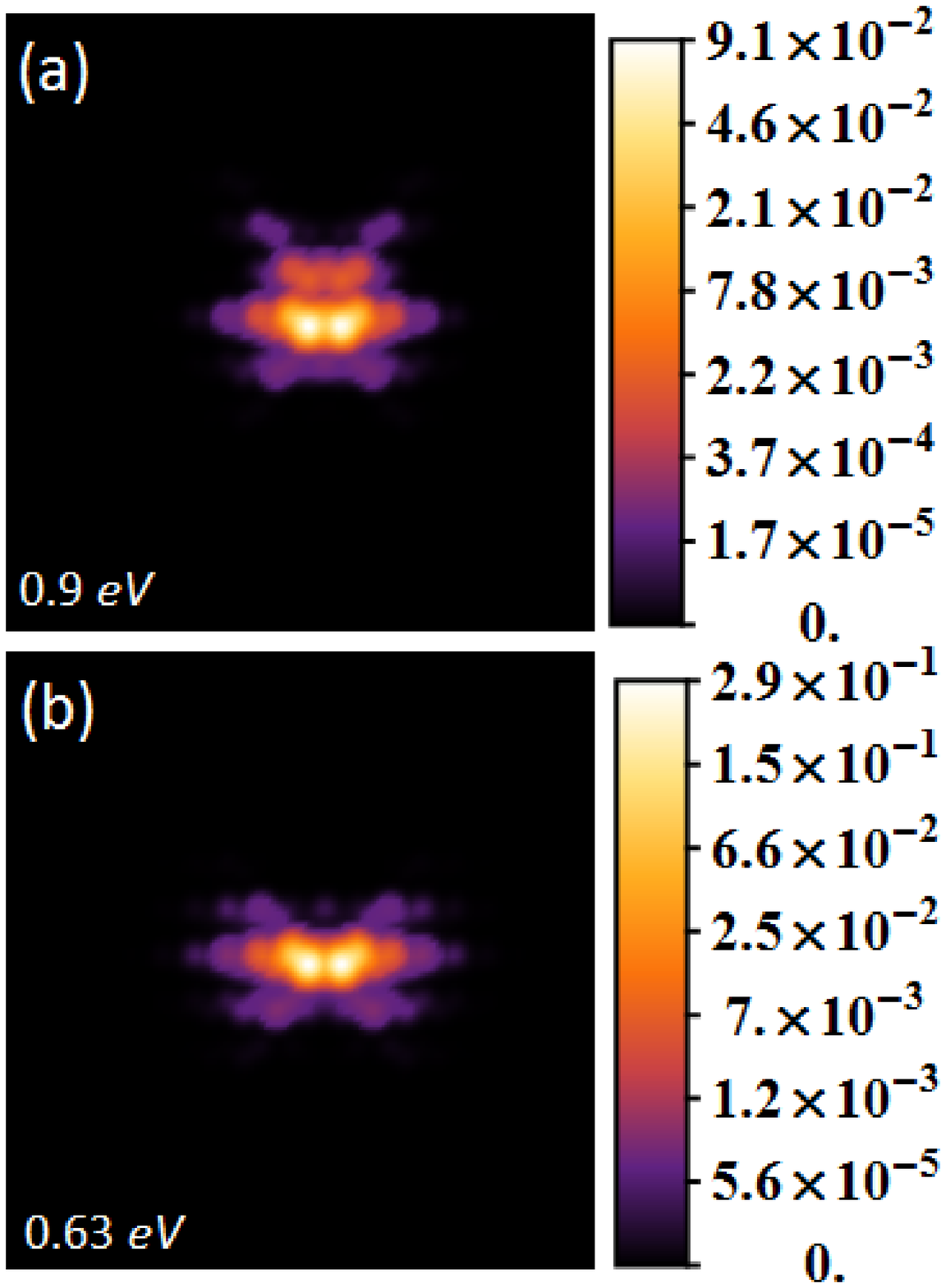}
\hspace{1.50mm}
\includegraphics[scale=0.5]{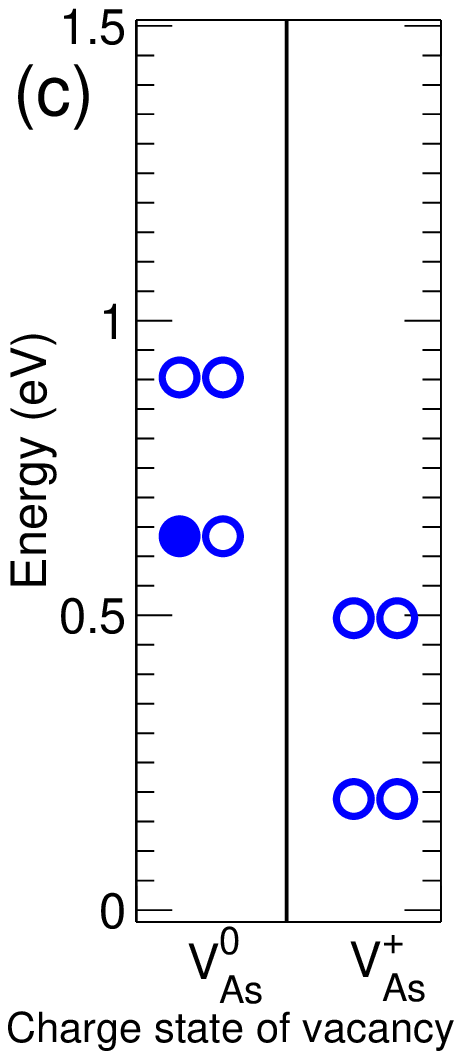}
\hspace{2.50mm}
\includegraphics[scale=0.32]{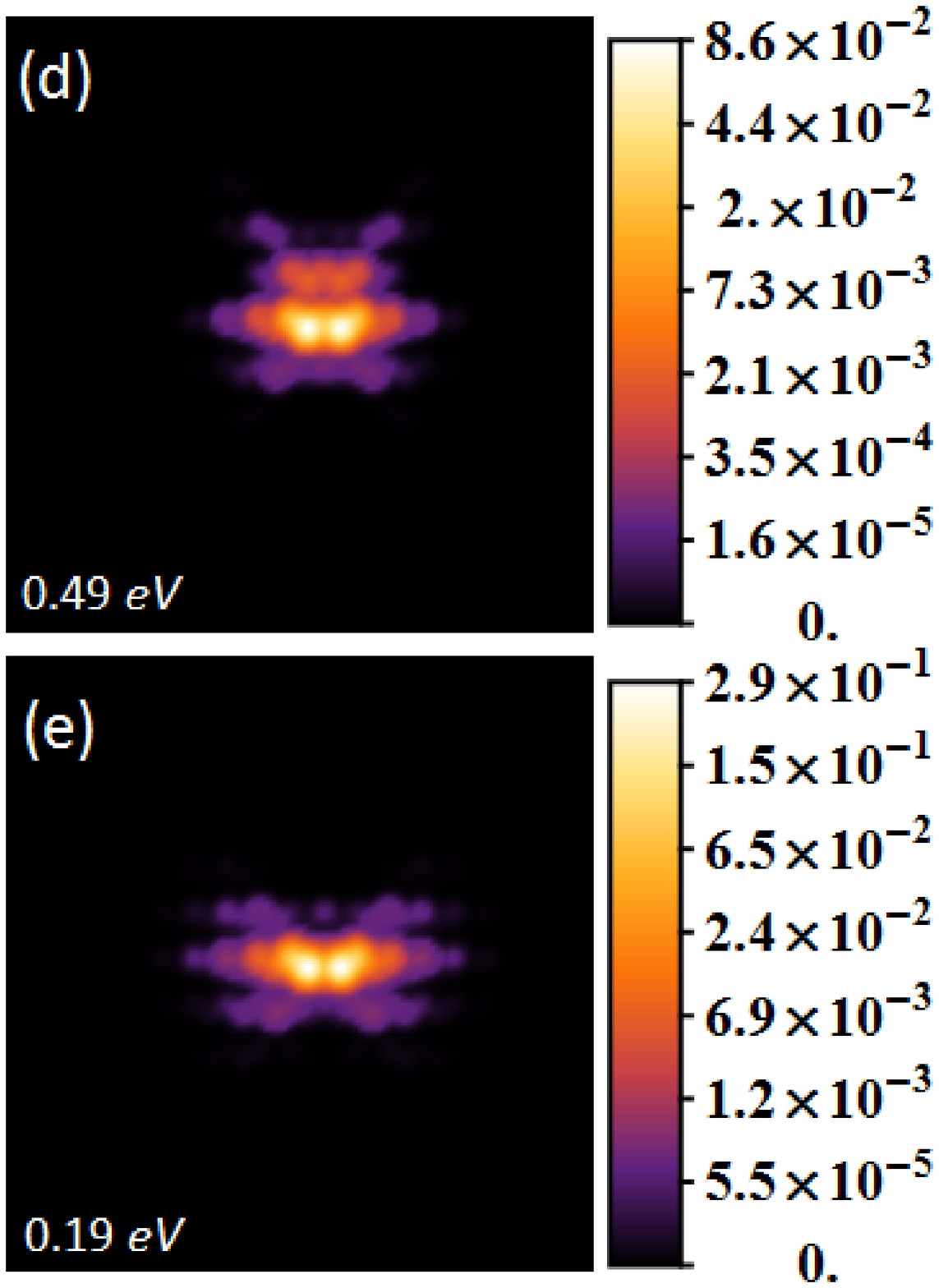}
\end{minipage}
		
\vspace{0.00mm}

\caption{(Color online) The level structure (c) and the local density of 
states (LDOS) (a-b, d-e) for an individual As
vacancy on the GaAs (110) surface. (a) and (b) LDOS of the two ${\rm V}_{As}^0$ bound
impurity levels in panel (c). The impurity levels are identified in (a) and (b) by their 
energies plotted on the left-hand side of panel (c). %I closed it here not Allan 
(c) The impurity structure for two different charge states of an As vacancy inside the GaAs gap. The zero
of energy corresponds to the valence band maximum of GaAs.
${\rm V}_{As}^0$ stands for neutral and ${\rm V}_{As}^+$ for positively charged As vacancies.
(d) and (e) LDOS of the two  ${\rm V}_{As}^+$ impurity levels in panel c).  The orbitals in
(d) and (e) are identified by their energies plotted on the right-hand side of panel (c).  
In each case the LDOS of the two orbitals in the degenerate pairs are identical.}
\label{fig:As_vac}
\end{figure*}
This section is divided into three sub-sections. We discuss the electronic properties of
individual As vacancies on the GaAs (110) surface in their neutral and positive charge states in Sec.~\ref{As_vac}.
In Sec.~\ref{As_vac_Mn}, we explain the effects of an As vacancy, in two different charge states,
on the electronic properties of one Mn impurity on the GaAs (110) 
surface. Finally in the last subsection, Sec.~\ref{As_vac_2Mn},
we analyze the electronic and magnetic properties of a pair of ferromagnetically
coupled Mn impurities in the presence of an As vacancy.
%
%
%%%%%%%%%%%%%%%%%%%%%%%%%%%%%%%%%%%%%%%%%%%%%%%%%%%%%%%%%%%%%%%%%%%%%%%%%%%%%%%%%
%
%
\subsection{Individual As vacancies on the GaAs (110) surface}
\label{As_vac}
Before investigating the effects of As vacancies
on the electronic and magnetic properties of nearby Mn impurities on the GaAs (110) surface,
we discuss the electronic properties of the As vacancies alone.
The $+1$ state, ${\rm V}_{As}^+$, is believed to be
the most stable charge state of an As vacancy on a $p-$type GaAs (110) 
surface~\cite{zhang_PRL_1996,chao_prb96, chelikowsky_prl96, gupta_science_2010}.
Therefore, throughout the paper, we only present results for
${\rm V}_{As}^+$ and the neutral state ${\rm V}_{As}^0$.
It is possible to switch reversibly between these two charge states 
using an STM tip~\cite{gupta_science_2010}.\\
Figure~\ref{fig:As_vac}(c) shows the level structure of ${\rm V}_{As}^0$ and ${\rm V}_{As}^+$
inside the GaAs gap. According to Fig.~\ref{fig:As_vac}(c), 
a ${\rm V}_{As}^0$ defect introduces two doubly-degenerate 
impurity levels in the gap. (It is important to point out that in our finite-cluster calculations
for a pure GaAs cluster, that is, without As vacancies, 
there are no levels in the gap. Therefore the levels plotted in Fig~\ref{fig:As_vac}(c) are directly due to the presence of the
As vacancy.) The energies of these levels are respectively at 0.63~eV (half-filled) 
and 0.9~eV (unoccupied) above the top of the valence band.  The position, degeneracy and  occupancy of these levels are
consistent with previous work based on first-principles calculations~\cite{Cheli_surf.sci_98}, 
in which these two levels are reported to be at 0.67~eV and 0.82~eV, respectively.

In our calculations the occupancy of these single-particle levels in the many-body ground state 
single Slater determinant
is determined simply by filling 
all electronic energies obtained by diagonalizing the Hamiltonian, 
starting
from the lowest levels,
with the valence electrons available in the finite
cluster. The electron number is given by 
\begin{equation}
N_e = 3\times N_{\rm Ga} + 5\times N_{\rm As}.
\label{electron_counting}
\end{equation}
For example, in a GaAs cluster of 8840 atoms with one As vacancy, the number of Ga atoms is $4420$ and the number of
As atoms is $(4420 - 1)$, leading to 35355 valence electrons in the cluster. (For a pure GaAs cluster
in the absence of the vacancy, this simple counting correctly reproduces, within a discrete particle-level picture, 
the separation between occupied valence band states and empty conduction band states.)\\ 
Panels~\ref{fig:As_vac}(a) 
and (b) show the local density of states (LDOS) for the
two ${\rm V}_{As}^0$ levels of Fig~\ref{fig:As_vac}(c). 
Note that the LDOS is plotted for the wavefunction of one of each 
doubly-degenerate level due to the similarity of the two.
The LDOS shows that the two-fold degenerate 
level at 0.63~eV is mainly localized on the two Ga atoms
neighboring the As vacancy. In fact 29$\%$ of the spectral weight 
is located on each neighboring Ga site. On the other hand, the LDOS for 
the level at higher energy (0.9~eV) shows less 
concentration of the spectral weight (only around 10$\%$) 
on the two neighboring Ga atoms.\\
By changing the charge state of the vacancy from ${\rm V}_{As}^0$ to 
${\rm V}_{As}^+$, the energies of the levels inside the gap changes 
as illustrated in Fig.~\ref{fig:As_vac}(c). In ${\rm V}_{As}^+$,
the two doubly-degenerate levels in the gap are 
pushed closer to the valance band than in the ${\rm V}_{As}^0$ case. Their positions are
at 0.19~eV and 0.49~eV. The level corresponding to the half-filled level in ${\rm V}_{As}^0$, is now unoccupied;
indeed, the simple electron counting of Eq.~\ref{electron_counting} gives one fewer electron 
because the vacancy has one positive charge.
These levels correspond to two levels at 0.27~eV and 0.43~eV, reported
previously~\cite{Cheli_surf.sci_98} for ${\rm V}_{As}^+$.
In spite of the energy shifts, the LDOS for these two levels, illustrated in  
Figs.~\ref{fig:As_vac}(d) and (e), resemble those of ${\rm V}_{As}^0$ case.\\
\begin{figure}[htp]
\includegraphics[scale=0.3]{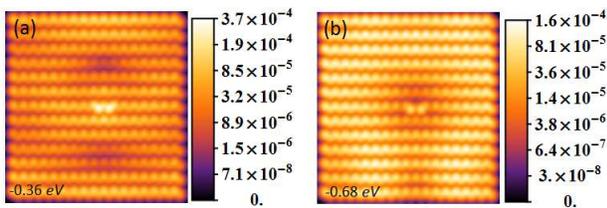}
\caption{(Color online) LDOS of As vacancy resonances in the valance band.
(a) LDOS for a neutral As vacancy level at -0.36~eV; (b) LDOS
for a positively charged vacancy at -0.68~eV.}
\label{fig:val-st}
\end{figure}
Before closing this section we should mention that both ${\rm V}_{As}^0$ and ${\rm V}_{As}^+$ 
have resonances in the valance band
that lie at -0.36~eV and -0.68~eV, respectively. 
These two energy values are close
to ones previously reported in Ref.~\onlinecite{Cheli_surf.sci_98}, -$0.32$~eV and $-0.66$~eV.
The LDOSs at these energies is plotted in Fig.~\ref{fig:val-st}.
Although the shape of the LDOS around the vacancy 
site is similar to the one of the in-gap states discussed above, the eigenfunctions are considerably delocalized,
as expected. 
We will not address these valance states in the rest of the paper;
instead we will focus on the level structure inside the GaAs gap.\\
%
%
%%%%%%%%%%%%%%%%%%%%%%%%%%%%%%%%%%%%%%%%%%%%%%%%%%%%%%%%%%%%%%%%%%%%%%%%%%%%%%
%
%
\subsection{One Mn impurity with nearby As vacancies}
\label{As_vac_Mn}
\begin{figure}[htp]
\centering
\includegraphics[scale=0.4]{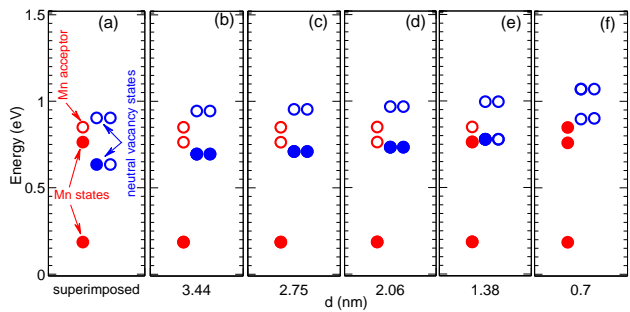}
\caption{(Color online) The level structure of the 
Mn-impurity/neutral-As-vacancy defect complex on the GaAs (110) surface for 
several different Mn-${\rm V}_{\rm As}^{0}$ separations.
(a) The superimposed level spectrum for decoupled Mn and 
neutral As vacancy defects.
(b)-(f) The level structure inside the GaAs 
gap for different impurity-vacancy separations shown on the bottom of each panel.
The blue circles indicates vacancy states and the red circles Mn states. 
Filled and empty circles are occupied and empty states, respectively.}
\label{fig:1Mn-n}
\end{figure}

We now consider the effects of ${\rm V}_{\rm As}^0$ and ${\rm V}_{\rm As}^+$ 
on the electronic properties of individual Mn dopants.
In the absence of As vacancies, 
a substitutional Mn atom (replacing a Ga atom) on the GaAs (110) surface introduces three 
levels in the GaAs gap. These three levels are shown in Fig.~\ref{fig:1Mn-n}(a) (red circles). 
The highest level in energy
is unoccupied (empty red circle) and, according to estimates based on STM experiments~\cite{yazdani_nat06,gupta_science_2010}, is a deep
level that  
lies at $\approx 0.85$~eV above the valence-band edge (${\rm Energy} =0$).
We will refer to this level as the {\it Mn-acceptor}.
The electronic and magnetic properties of the Mn-acceptor on the GaAs (110) surface
have been studied extensively in the last 
eight years~\cite{yazdani_nat06,Jancu_PRL_08, scm_MnGaAs_paper1_prb09, 
gupta_science_2010, garleff_prb_2010,mc_MF_2013, PhysRevB.89.165408}. 
The model described by Eq.~\ref{hamiltonian}
reproduces essentially all the features seen in STM experiments.\\
In Fig.~\ref{fig:1Mn-n} we plot the electronic structure inside the GaAs gap for the
${\rm V}_{\rm As}^0$--Mn-impurity system for several different impurity-vacancy separations. 
In panel (a) of this figure
we superimpose the Mn-impurity states, calculated in the absence of any vacancy,
onto the ${\rm V}_{\rm As}^0$ states, calculated in the absence of the Mn impurity.
Blue color indicates states localized near the vacancy and red color indicates a state localized 
near the Mn, while filled and empty circles distinguish states that are occupied and empty
in an independent electron approximation.  
This panel will be used as a reference case to compare
the electronic structures of the Mn impurity and the As vacancy
at different impurity-vacancy separations.
Note that in Fig.~\ref{fig:1Mn-n}(a), which describes the case of large impurity-vacancy separations,
the highest occupied Mn level is higher in energy than the lowest unoccupied 
${\rm V}_{\rm As}^0$) orbital.  Panels~\ref{fig:1Mn-n}(b)-(f) show how the Mn impurity and As vacancy states
are perturbed at various finite distances from one another.
As the distance between Mn impurity and As vacancy decreases,
the ${\rm V}_{As}^0$ states are slightly pushed up toward the conduction band, while
the Mn states, and more specifically the Mn-acceptor, defined as the Mn highest 
unoccupied level in panel~\ref{fig:1Mn-n}(a), remain unshifted.\\
In our model the reason for the increase of the vacancy energy levels is the presence of the one-body potential
in  Eq.~\ref{hamiltonian}, coming from the Mn ion. This potential is positive and repels electrons from
the Mn core -- and it has the effect of raising also the single-particle energy levels introduced by the
presence of the vacancy.  At larger separations, the long-range contribution of this potential 
(last but one term in Eq.~\ref{hamiltonian})
is mainly responsible for this effect, as one can evince from the slow increase of the vacancy levels with
decreasing distance $d$. However, in close proximity of the Mn impurity ( Fig.~(f)), short range
contributions coming from $V_{\rm Corr}$ can indirectly affect the vacancy levels, and cause a
more pronounced increase of the energy levels at the shortest separation $d= 0.7$ nm.\\ 
In an independent electron approximation, the variation of
the energy levels of ${\rm V}_{As}^0$ with respect to the Mn  levels,
causes the occupancy of both ${\rm V}_{As}^0$ and Mn levels to depend on the Mn-${\rm V}_{As}^0$ separation.
%First of all, the electron counting of Eq.~\ref{electron_counting} must now be modified by the presence of the Mn impurity, 
%which contributes
%two electrons instead of the three of Ga.
Fig.~\ref{fig:1Mn-n}(b)-(f) shows that for separations larger than  $\approx 1.4$ nm, 
the lowest of the two (doubly-degenerate)
${\rm V}_{As}^0$ levels is lower than the second Mn level (the one just below the Mn ``acceptor''). 
In an independent electron approximation,
occupation of the energy levels of the two separate defects, shown in panel (a),
would therefore be modified by having the ${\rm V}_{As}^0$ level doubly occupied and the Mn level empty. 
This change in occupation would therefore make the ${\rm V}_{As}$ negatively charged and the 
Mn acceptor doubly charged.  Clearly this result cannot be taken literally. 
In fact, double-occupancy by electrons of the lowest of the two ${\rm V}_{As}^0$ levels
is very unfavorable energetically when electron-electron correlations are included, 
We expect that fluctuations in the 
charge states near the different defects to play a role in the 
exact many-body ground state.  It is quite possible that the mean electron occupations of the
two defect energy levels remains close to the one of panel (a) at all separations, 
with the possible exception of the shortest distance at
$d= 0.7$ nm; see panel (f).\\
\begin{figure}[htp]
\begin{minipage}[h]{1.0\linewidth}
\centering
\includegraphics[scale=0.3]{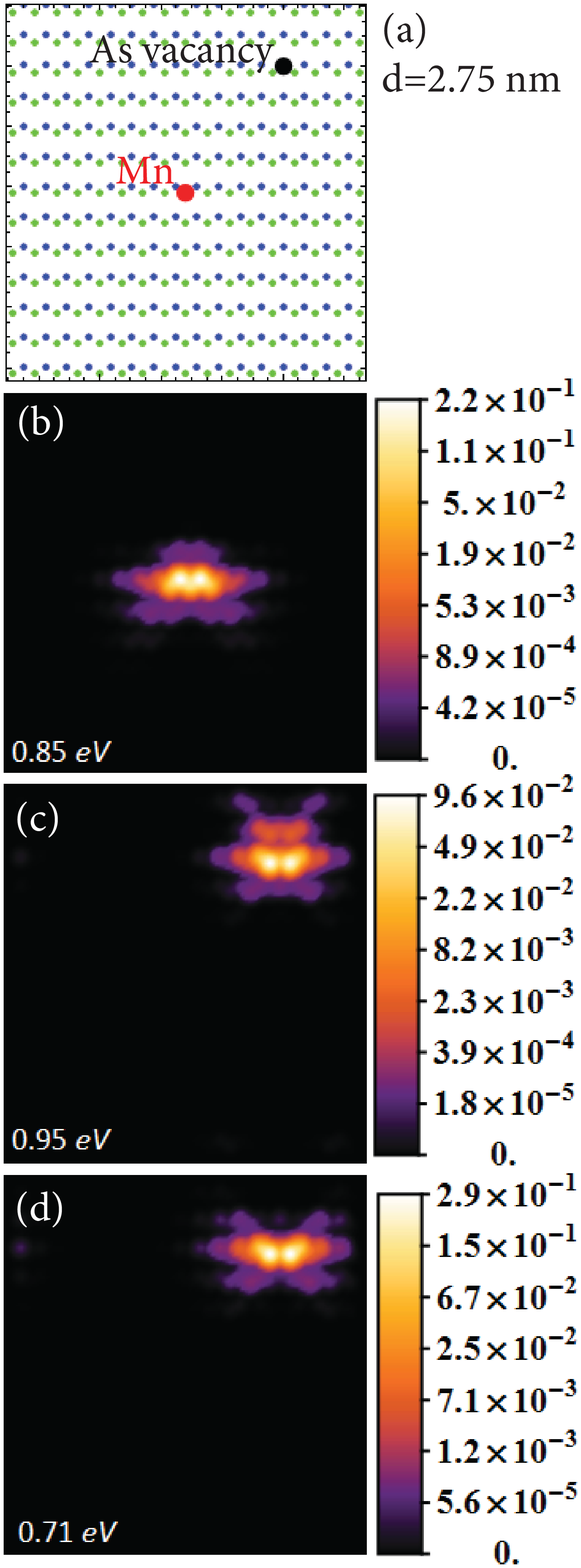}
\hspace{1.50mm}
\includegraphics[scale=0.3]{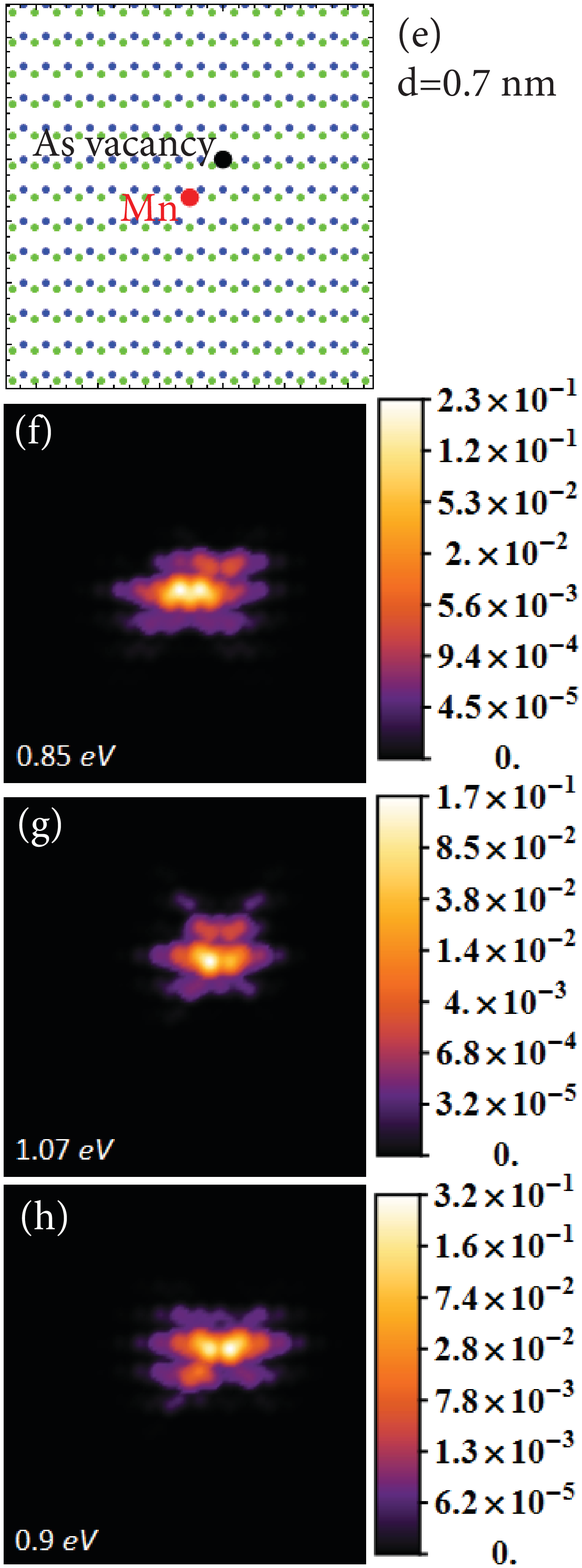}
\end{minipage}
		
\vspace{0.00mm}

\caption{(Color online) The LDOS of a Mn-acceptor and a neutral 
As vacancy on the GaAs (110) surface for two Mn-${\rm V}_{\rm As}^0$ separations. 
Panels (a) and (e) show the position of the Mn (red circle) and As 
vacancy (black circle) on the (110) surface. The three pictures below on the left and right refer to
these two situations.
(b) LDOS of the Mn-acceptor and (c)-(d) LDOS of two ${\rm V}_{\rm As}^0$
states for the level configuration shown in Fig~\ref{fig:1Mn-n}(c). 
(f)-(h) are similar to (b)-(d) for the configuration shown in Fig~\ref{fig:1Mn-n}(f).
The LDOS is plotted for one of the two ${\rm V}_{\rm As}^0$ states belonging to
each degenerate level.}
\label{fig:1Mn-n-LDOS}
\end{figure}
Figure~\ref{fig:1Mn-n-LDOS} shows the LDOS of the Mn-acceptor and  ${\rm V}_{\rm As}^0$ states
for the levels shown in panels~\ref{fig:1Mn-n}(c) and~\ref{fig:1Mn-n}(f).
The left column in Fig.~\ref{fig:1Mn-n-LDOS} refers to the case when the distance between the Mn
and ${\rm V}_{\rm As}^0$ is $2.75$~nm, while the right column is for $0.7$ nm.
Panels~\ref{fig:1Mn-n-LDOS}(b) and (f) are the LDOS of the Mn-acceptor
and panels~\ref{fig:1Mn-n-LDOS}(c), (d), (g) and (h) are the ${\rm V}_{\rm As}^0$ states.
The position of the Mn impurity and ${\rm V}_{\rm As}^0$ on the (110) surface 
for each column is shown on top of the column 
(see panels~\ref{fig:1Mn-n-LDOS}(a) and (e)).
The LDOSs on the left column are similar to the LDOSs
of the Mn and the ${\rm V}_{\rm As}^0$ in the absence of the other defect, 
due to the large distance between them.
When the vacancy is at $d = 0.7$ nm from the Mn (right column), the symmetry of both the Mn-acceptor
and the ${\rm V}_{\rm As}^0$ lowest state (which is closest in energy to the Mn levels)
is significantly modified.\\
\begin{figure}[htp]
\centering
\includegraphics[scale=0.4]{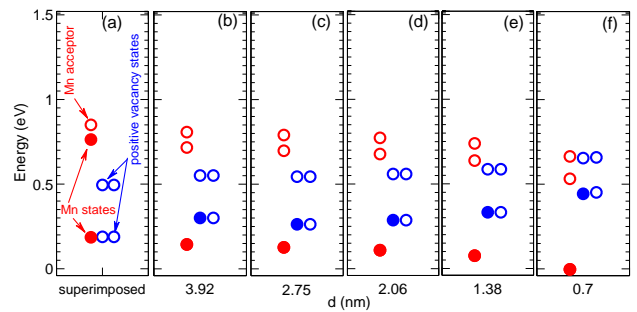}
\caption{(Color online) The level structure of a Mn and 
a positively charged As vacancy on a (110) GaAs 
surface for different separations, as in Fig.~\ref{fig:1Mn-n}.
Blue colors indicate vacancy states and red colors indicate Mn states. 
Filled and empty circles distinguish states that are occupied and empty
in an independent electron approximation.}
\label{fig:1Mn-p}
\end{figure}
Let us now consider the effect of ${\rm V}_{\rm As}^+$ on 
the Mn-acceptor on the GaAs (110) surface. 
In Fig.\ref{fig:1Mn-p}~we plot the electronic structure for this system inside the GaAs energy gap. 
Panel~\ref{fig:1Mn-p}(a) shows the energy levels of the Mn-impurity and the ${\rm V}_{\rm As}^+$
calculated separately. Again, we will use this level structure, which can be viewed as representing the case
of a  ${\rm V}_{\rm As}^+$ placed at a very large separation from the  Mn impurity, 
as a reference for the coupled system.
Note that since now the As vacancy has a charge of +1, the total electronic counting given by 
Eq.~\ref{electron_counting} is decreased by one unit.
Panels~\ref{fig:1Mn-p}(b)-(f)
show how the energy levels for the Mn-${\rm V}_{\rm As}^+$ system change as ${\rm V}_{\rm As}^+$ is placed closer
to the Mn impurity.
The ${\rm V}_{\rm As}^+$ states are pushed up toward the conduction band
as in the case of ${\rm V}_{\rm As}^0$. % due the same mechanism used to explain Fig.~\ref{fig:1Mn-n},
Now however, in contrast to the ${\rm V}_{\rm As}^0$ case and in 
agreement with experiment~\cite{gupta_science_2010},  the energy of the Mn-acceptor state also 
moves toward the valance band.\\
Let us look at this result first from the experimental point of view.
We know that the bands in a semiconductor bend in the presence 
of nearby metallic gates, such as the metallic tip of a STM ~\cite{Feenstra_vst_87, Feenstra_vst_03, Ishida_prb_09}.
This phenomenon, known as tip-induced band-bending (TIBB), is important for the correct interpretation of STM data,
particularly in determining the position of impurity levels occurring inside the gap. 
In STM experiments the TIBB is ``negative'', that is, it causes a downward band-bending 
for all voltages below $V_{\rm FB}$, the voltage that ensures the flat-band condition, ${\rm TIBB}= 0$.
(In Ref.~\onlinecite{gupta_science_2010}, $V_{\rm FB} = + 1.6 {\rm V}$.)
The charged vacancies investigated in Ref.~\onlinecite{gupta_science_2010} and 
in the present paper can have a similar effect on the bands of the semiconductor host. 
In analogy to TIBB, this phenomenon is called vacancy-induced band-bending (VIBB).
VIBB from a positively charged As vacancy is expected to induce a further downward band-bending, in addition
to the one coming from the TIBB at ``negative'' voltages. 
In a rigid band-bending model, where the impurity levels follow rigidly the shift of the host bands, 
a downward band-bending of the valence band (caused by a combination of TIBB and VIBB) 
should result in a shift of the acceptor resonant structure,
measured in the differential conductance spectra
vs. the applied STM bias voltage. In particular, this model would predict that the Mn-acceptor resonance should 
shift toward higher voltages as ${\rm V}_{\rm As}^+$ is moved closer to the Mn\cite{gupta_science_2010}. 
This would lead to an apparent increase
of the acceptor binding energy. The careful experimental results of Ref.~\onlinecite{gupta_science_2010} find the opposite:
the Mn-acceptor binding energy is insensitive to varying TIBB conditions, indicating that the surface-layer Mn levels 
are detached from the GaAs valence
band. Furthermore, the Mn-acceptor binding energy, extracted from the
resonances in the differential conductance taken on the Mn,
decreases when ${\rm V}_{\rm As}^+$ is brought closer to the impurity.
By ruling out a rigid band-bending as the main influence of ${\rm V}_{\rm As}^+$ on the Mn-acceptor,
the authors suggested that the decrease in the binding energy of the Mn-acceptor is
the result of the direct Coulomb repulsion between the ${\rm V}_{\rm As}^+$ and the Mn-hole~\footnote{It is interesting to 
point out that in  Ref.~\onlinecite{gupta_science_2010} in-gap resonances of STM spectra taken on subsurfaces Zn acceptors
shift toward {\it higher} voltage as  ${\rm V}_{\rm As}^+$ is moved closer. A numerical simulation shows that in this
case the rigid band-bending model reproduces perfectly the influence of the As vacancy on the acceptor position.}.
In agreement with their experiment,
the $(+1)$-charged As vacancy in our calculations, 
acting as a positive point charge, repels
and delocalizes the charged hole bound to the Mn, causing a decrease of 
its binding energy. 
The effect makes it possible to locally manipulate the
electronic properties of Mn dopants in GaAs by means of an electric field 
produced by a nearby ${\rm V}_{\rm As}^+$.\\
\begin{figure}[htp]
\begin{minipage}[h]{1.0\linewidth}
\centering
\includegraphics[scale=0.3]{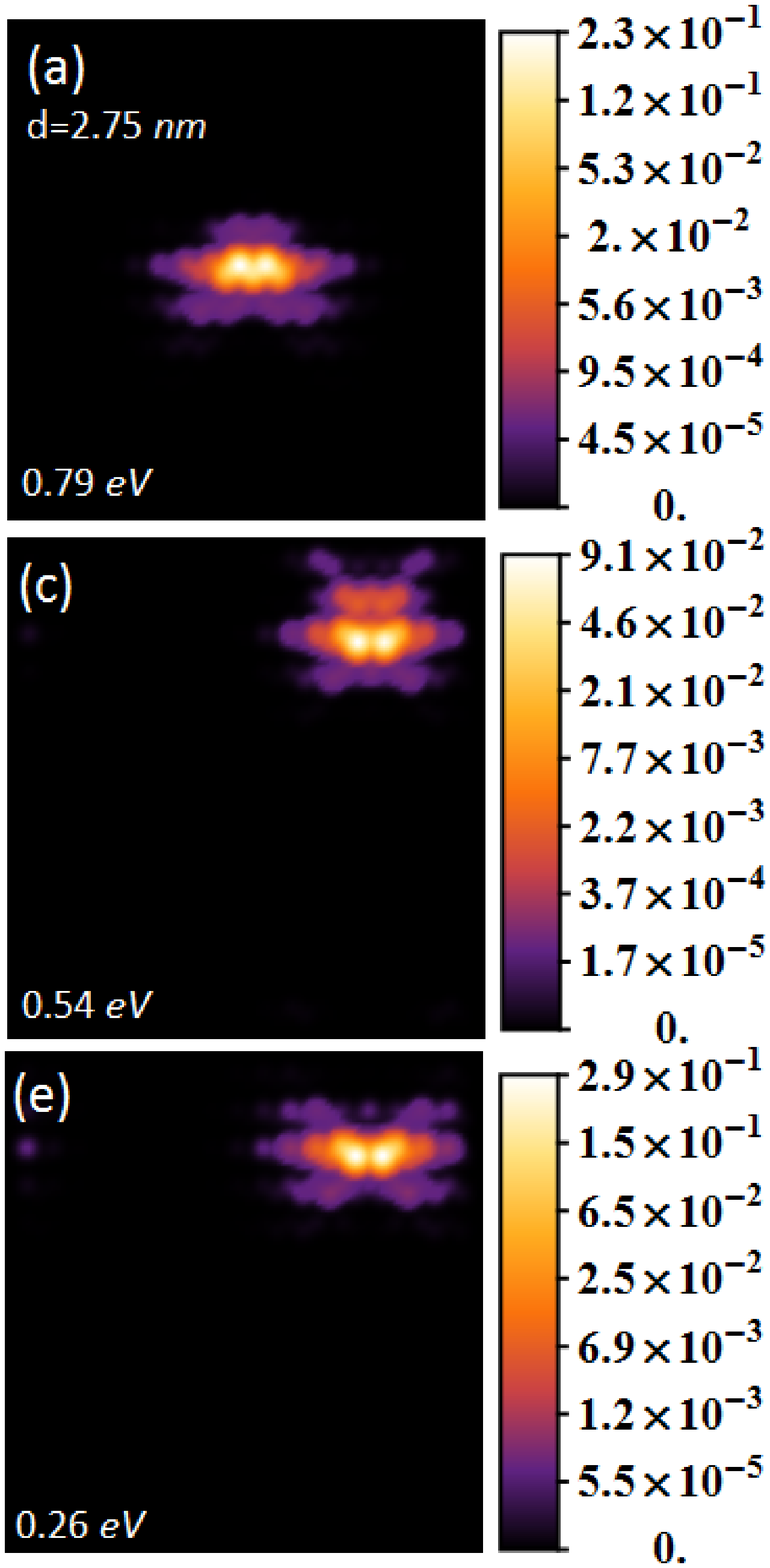}
\hspace{1.50mm}
\includegraphics[scale=0.3]{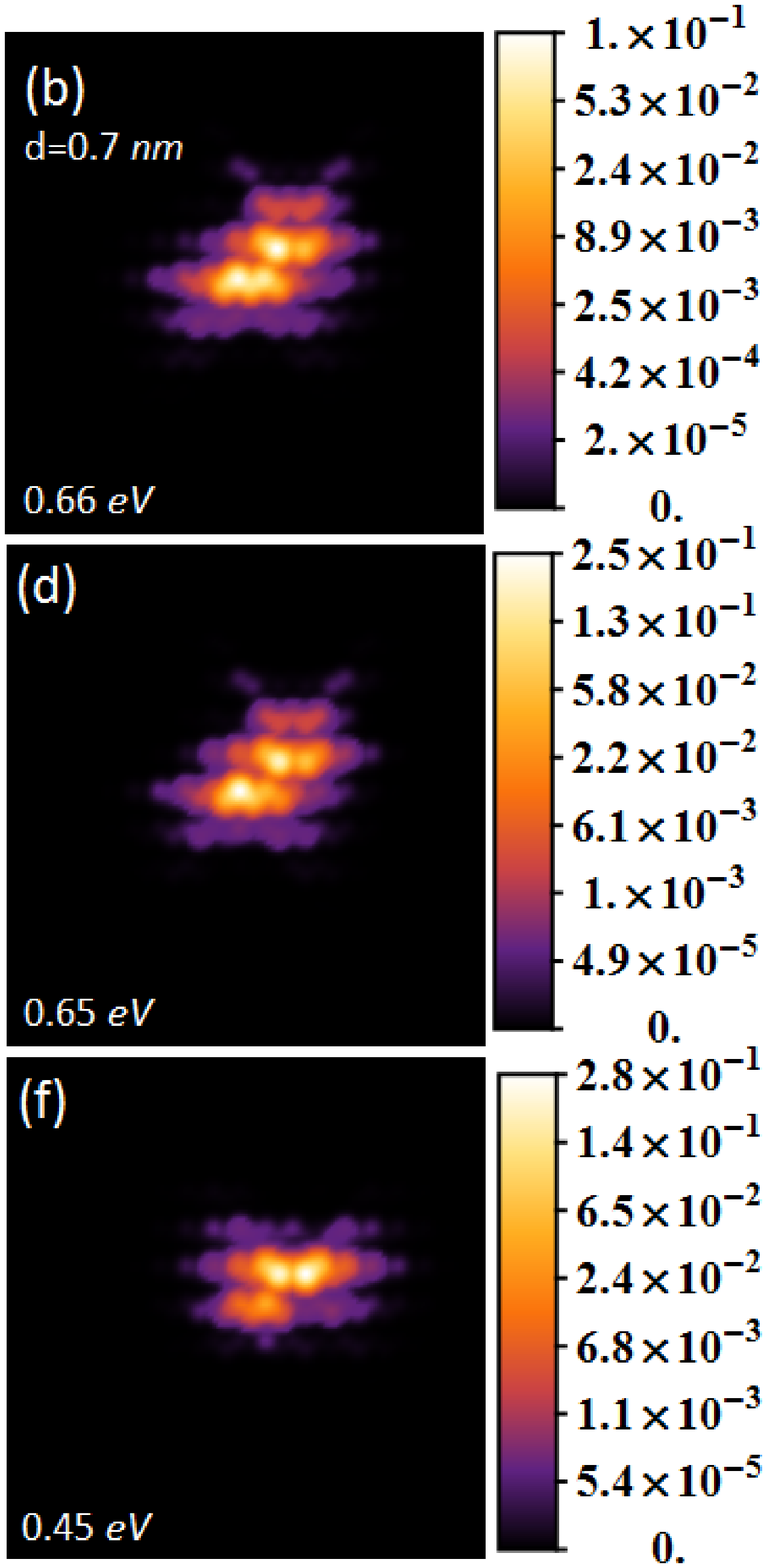}
\end{minipage}
		
\vspace{0.00mm}

\caption{(Color online) LDOS of the Mn-acceptor state and a ${\rm V}_{\rm As}^+$ state
on the GaAs (110) surface for two Mn-${\rm V}_{\rm As}^+$ separations $d$. 
Left column: $d=2.75$ nm: right column: $d=0.7$ nm.
Positions of the Mn impurity and the As 
vacancy on the GaAs surface are the same of Fig.~\ref{fig:1Mn-n-LDOS}.
The three LDOSs of the left and right column correspond 
to the energy level configurations given in Fig.~\ref{fig:1Mn-p}(c)
and~\ref{fig:1Mn-p}(f), respectively. (a) and (b) represent Mn-acceptor states, while
(c)-(e) and (d)-(f) represent ${\rm V}_{\rm As}^+$ states 
The two states of each degenerate level of ${\rm V}_{\rm As}^+$
have quite similar LDOSs, except for the LDOS of level at $0.65$ eV,
shown in panel (d), where we plot the combined LDOSs. 
The two separate LDOSs for this panel are plotted in Fig~\ref{fig:1Mn0.7p-spl}.}
\label{fig:1Mn-p-LDOS}
\end{figure}
Figure~\ref{fig:1Mn-p-LDOS} shows the LDOS for the mid-gap states of the
Mn-impurity--${\rm V}_{\rm As}^+$ complex, for the same two defect separations, $d= 2.75$ (left column) and
$d= 0.7$~nm (right column),
already considered in Fig.~\ref{fig:1Mn-n-LDOS}
for the case of the Mn-impurity--${\rm V}_{\rm As}^0$ complex. 
The states in this picture
correspond to the energy levels shown in Figs.~\ref{fig:1Mn-p}(c) and~\ref{fig:1Mn-p}(f).
The two top panels (a)-(b) are the LDOSs for the Mn-acceptor state (that is, the highest unoccupied Mn level 
in Figs.~\ref{fig:1Mn-p}(c) and~\ref{fig:1Mn-p}(f).) 
A careful analysis of these LDOSs reveals that the Mn-acceptor wavefunction 
becomes less localized around Mn impurity as the positive As vacancy is brought closer to the Mn.
For example, for $=2.75$~nm 23$\%$ of the acceptor spectral weight resides on the Mn. 
This value decreases to only 10$\%$ when the separation decreases to $d= 0.7$~nm.
This result is consistent with and further confirms the physical mechanism put forward above to account for the
decrease of the Mn-acceptor binding energy with decreasing vacancy separation: a positively charged As vacancy
repels the Mn-acceptor and therefore delocalizes the bound acceptor, leading to a decrease of its binding energy.\\
Panels (c)-(d) and (e)-(f) in Fig.~\ref{fig:1Mn-p-LDOS}
refer to the two doubly-degenerate ${\rm V}_{\rm As}^+$ levels. As for the neutral As vacancy, 
in all cases except for panel (d)
the two states belonging to the same doubly-degenerate level have very similar LDOS, and therefore only one
of them is shown in the figure. The situation in  Fig.~\ref{fig:1Mn-p-LDOS}(d) is special. 
This LDOS is in fact the sum of
the LDOSs for the two states corresponding to the As energy level located at $\approx 0.65$~eV. 
(See Fig.~\ref{fig:1Mn-p}(f).) 
A careful analysis shows that the energies of these two states are in fact split by $\simeq 5$~meV.
In this case the two wavefunctions are also quite different, as shown in Fig.~\ref{fig:1Mn0.7p-spl}.
One of the two states [Fig.~\ref{fig:1Mn0.7p-spl}(a)] is mainly located on one of the neighboring Ga atoms, 
while the other [Fig.~\ref{fig:1Mn0.7p-spl}(b)] has a
significant contribution also on the As atoms close to the Mn. Clearly, at this close separation
the Mn-dopant acts like an impurity potential for the As doublet which is therefore slightly split. In fact,
the three top states in Fig.~\ref{fig:1Mn-p}(f) (one Mn-impurity and two ${\rm V}_{\rm As}^+$) 
are all quite close in energy, and their corresponding wavefunctions are
strongly hybridized.
Note that the combined LDOS of these two ${\rm V}_{\rm As}^+$ states 
resembles quite closely the LDOS of the Mn-acceptor.\\  
\begin{figure}[htp]
\includegraphics[scale=0.3]{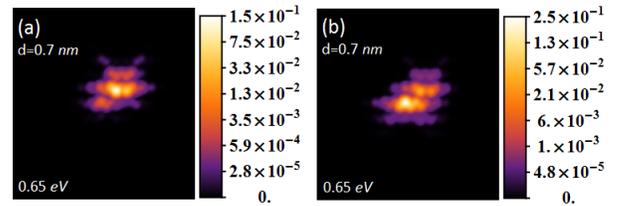}
\caption{(Color online) The energy resolved LDOS for two states 
of a positively charged As vacancy at 0.65~eV in Fig~\ref{fig:1Mn-p}(f).}
\label{fig:1Mn0.7p-spl}
\end{figure}
\begin{figure}[htp]
\centering
\includegraphics[scale=0.54]{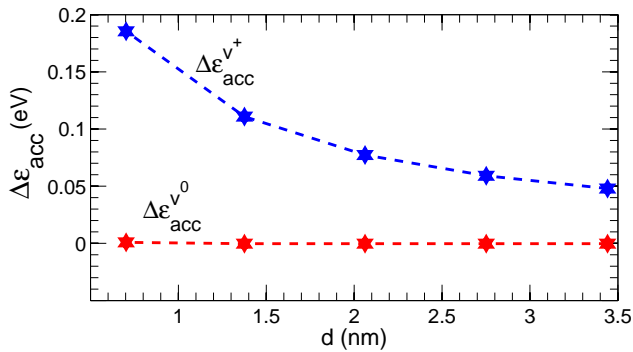}
\caption{(Color online)  Change in the binding energy of the Mn-acceptor 
caused by a neutral and positive As vacancy (Eq.~\ref{acc-ch}).
Blue and red asterisks represent the change caused by a positively 
charged and a neutral As vacancy, respectively.}
\label{fig:Acceptor}
\end{figure}
We conclude this section by quantifying the change in the Mn-acceptor binding energy
as a function of Mn-impurity--${\rm V}_{\rm As}$ separation for the two cases of a neutral and charged vacancy.
We define the shift in the Mn-acceptor binding-energy caused by the As vacancy,
$\Delta \epsilon_{\rm acc}$ as [see Eq.~\ref{acc-ch}], 
\begin{align}
\Delta \epsilon_{\rm acc}(d) =\epsilon_{\rm acc}^{\rm ref}-\epsilon_{\rm acc}^{{\rm V}_{\rm As}^{0/+}}(d)\;,
\label{acc-ch}
\end{align}
where $\epsilon_{\rm acc}^{\rm ref}$ is the Mn-acceptor binding energy in
the absence of the As vacancy, (reference case), and $\epsilon_{\rm acc}^{{\rm V}_{\rm As}^{0/+}}(d)$ 
is the Mn-acceptor binding energy
when ${\rm V}_{\rm As}^0$/${\rm V}_{\rm As}^+$ is at the distance $d$ from the Mn impurity.
As usual, in all cases we define the Mn-acceptor binding energy as the energy of the highest unoccupied
energy level of the Mn-impurity in the GaAs energy gap~\footnote{For the case of one Mn on the (110)
surface, the Mn-acceptor 
defined in this way is always a mid-gap state. For a Mn pair discussed in the next section, we will see
that one of the two acceptors ends in the conduction band.}.
Note that a positive value of $\Delta \epsilon_{\rm acc}$ corresponds to a decrease in binding energy.\\
In Fig.~\ref{fig:Acceptor} we plot $\Delta \epsilon_{\rm acc}(d)$ vs. $d$ for both
the Mn-impurity-${\rm V}_{\rm As}^0$ and  Mn-impurity--${\rm V}_{\rm As}^+$ system.
This figure summarizes our numerical results for these two cases. 
${\rm V}_{\rm As}^0$ does not alter the Mn-acceptor binding energy even 
at the shortest impurity-vacancy separations (red asterisks in the figure).
In contrast the local electric field generated by ${\rm V}_{\rm As}^+$  
has the effect of decreasing $\epsilon_{\rm acc}^{{\rm V}_{\rm As}^{+}}$,
as the As vacancy is brought closer to Mn.\\
As shown in Fig.~\ref{fig:Acceptor}, the decrease in the Mn-acceptor binding energy
with respect to the reference value
is 0.11~eV at a separation $d= 1.38$~nm, which matches reasonably well to the
experimental value of $\approx$0.16~eV measured at a separation $d= 1.42$~nm~\cite{gupta_science_2010}.
At a separation  $d= 0.7$~nm the calculated acceptor binding energy drops further 
down by 0.19~eV. There is at present
no experimental value at this close separation to compare with.\\
%cmc: 
%{\color{red} cmc: what is the experimental value of the binding-energy decrease at this separation?\\}
%rm
%{\color{blue} rm: We don't know, the shortest separation in Gupta's paper is 1.42 nm\\}
%%%%%%%%%%%%%%%%%%%%%%%%%%%%%%%%%%%%%%%%%%%%%%%%%%%%%%%%%%%%%%%%%%%%%%%%%%%%%%
%
%
%
\subsection{Pairs of ferromagnetically 
coupled Mn impurities with nearby As vacancies}
\label{As_vac_2Mn}
\begin{figure}[htp]
\centering
\includegraphics[scale=0.4]{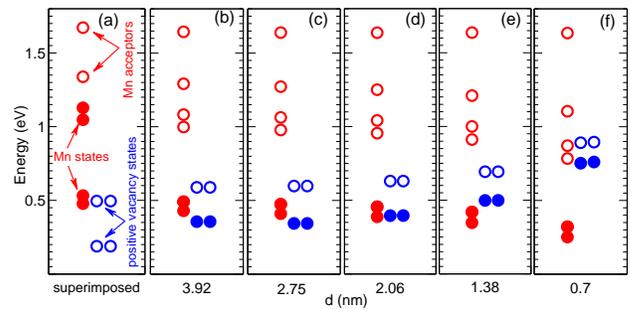}
\caption{(Color online) The level structure of a pair of ferromagnetically coupled Mn impurities and 
a positively charged As vacancy on the GaAs (110) surface, for different defect separations; 
see the caption of Fig.~\ref{fig:1Mn-n}.
Blue color indicates vacancy states and red color Mn states. 
Filled and empty circles indicate occupied and empty states levels in an 
independent-electron approximation.}
\label{fig:2Mn-p}
\end{figure}

We will now discuss the influence of ${\rm V}_{\rm As}^0$ and ${\rm V}_{\rm As}^+$
on a pair of ferromagnetically coupled Mn impurities on the (110) surface of
GaAs.\\ 
%cmc: 
%{\color{red} cmc: There are no comments on how the Mn pair is constructed. What distance, which symmetry direction.
%      Please check what I wrote and fill in details.\\}
%rm
%{\color{blue} rm: I completed it as following\\}
In all our calculations we will keep the distance between the two Mn impurities fixed, and we will vary 
the distance between the Mn pair and the As vacancy.
Specifically, the two Mn dopants are chosen to substitute two nearest-neighbor 
Ga atoms along the $<110>$ symmetry direction on the $(110)$ surface. Their separation is $0.4$ nm.
At this distance the exchange-coupling strength between their spins is maximum\cite{scm_MnGaAs_paper2_prb2010}.\\
\begin{figure}[htp]
\centering
\includegraphics[scale=0.4]{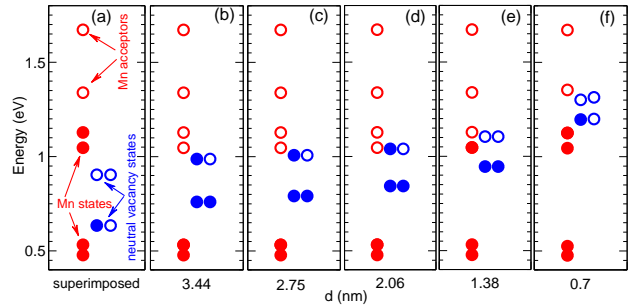}
\caption{(Color online) The level structure of a pair of 
ferromagnetically coupled Mn impurities and 
a neutral As vacancy on the GaAs (110) surface for different separations; 
see the caption of Fig.~\ref{fig:1Mn-n}.
Blue color indicates vacancy states and red color Mn states. 
Filled and empty circles are occupied and empty states, respectively.}
\label{fig:2Mn-n}
\end{figure}
\begin{figure}[htp]
\centering
\includegraphics[scale=0.6]{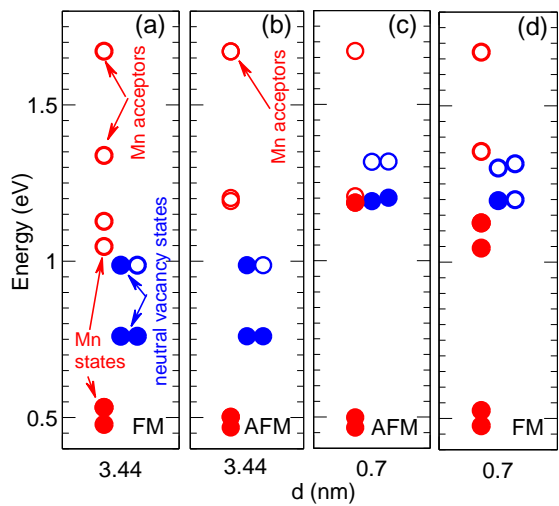}
\caption{(Color online) The level structure of a pair of Mn impurities and 
a neutral As vacancy on the GaAs (110) surface for two different separations; 
see the caption of Fig.~\ref{fig:1Mn-n}.
(a) and (d) pair of Mn are ferromagnetically (FM) coupled,
(b) and (c) pair of Mn are antiferromagnetically (AFM) coupled.
Blue color indicates vacancy states and red color Mn states. 
Filled and empty circles are occupied and empty states, respectively.
The topmost Mn-acceptor level for the AFM case is doubly-degenerate.}
\label{fig:2Mn-n-FM-AFM}
\end{figure}
Figure~\ref{fig:2Mn-n} shows the energy level structure of a ${\rm V}_{\rm As}^0$ and a pair of 
ferromagnetically coupled Mn impurities inside the GaAs gap. 
As a reference, in Fig.~\ref{fig:2Mn-n}(a)
we superimposed the level structure of a ${\rm V}_{\rm As}^0$ 
as well as the topmost six levels of a pair of 
Mn impurities inside the GaAs gap, calculated separately.
For a pair of Mn impurities the two topmost levels of these six levels occurring
in the GaAs gap are unoccupied.
They are the two acceptor levels introduced by the two Mn impurities, located at 1.34~eV and 1.67~eV,
respectively~\footnote{Note 
that for the choice of the parameters presently used in the
TB Hamiltonian,
the topmost Mn-acceptor level with energy equal to 1.67 eV, is in fact merging with the conduction band. 
Experimentally this level is found to be at 1.4 eV, that is, 
still inside the band gap but quite close to the conduction band edge (located at 1.52 eV). 
Its position in our
calculations can be tuned to the experimental value 
by choosing an appropriate value of $V_{\rm Corr}$.\cite{scm_MnGaAs_paper2_prb2010}}. 
The energies  of the two acceptor levels are split by a few hundreds of meV only when
the spins of the two Mn are coupled ferromagnetically -- which is usually the configuration of
lowest energy~\cite{yazdani_nat06, tang_spie2009, scm_MnGaAs_paper2_prb2010}.
In contrast, as shown in Fig.~\ref{fig:2Mn-n}(a), if the Mn pair is coupled antiferromagnetically, 
there is no energy splitting between the two
unoccupied acceptors: the topmost unoccupied level is doubly-degenerate. 
Therefore, the splitting between acceptors  for a ferromagnetically coupled pair is somewhat 
related to the ferromagnetic exchange coupling between them\cite{yazdani_nat06,tang_spie2009, scm_MnGaAs_paper2_prb2010}. 
Since these splittings can be directly probed in STM experiments,
it is possible to probe the exchange coupling through tunneling spectroscopy via the acceptor states.
Furthermore, since the acceptor energy levels can be electrically manipulated by nearby defects such as As vacancies, 
the exchange coupling can be
indirectly modified by the presence of these extra defects.
In Figs.~\ref{fig:2Mn-n}(b)-(f) one can see how the energy levels and in particular how the splitting between the Mn-acceptors
is modified by the presence of a ${\rm V}_{\rm As}^0$
placed at successively shorter separations from the Mn pair. 
As for the case of an individual Mn, we find that the ${\rm V}_{\rm As}^0$ levels are typically pushed up in energy
toward the conduction band when approaching the Mn. However, the levels associated with the Mn impurity are hardly modified;
in particular the acceptor splitting remains constant with the exception of the shortest separation $d= 0.7$ nm, where it decreases slightly. 
As shown in the figure, 
the electron occupancy of both  ${\rm V}_{\rm As}^0$ levels and some of the lower
Mn levels change as the ${\rm V}_{\rm As}^0$ is brought closer to the Mn pair. As mentioned in Sec.~\ref{As_vac_Mn}, some of these results
might be affected by on-site electron correlations not included in our model. Note, however, that the acceptor states remain unoccupied at
all separations that we considered. It is important to point out that the acceptor splitting is nonzero only for a ferromagnetically coupled
Mn pair even in the presence of the ${\rm V}_{\rm As}^0$. This can be clearly seen in Fig.~\ref{fig:2Mn-n-FM-AFM}, 
showing the energy levels at two separations
for both a ferromagnetically (FM) coupled Mn pair as well as for 
an antiferromagnetically (AFM) coupled one. In the second case, the acceptor splitting is
identically zero; the topmost level (located at 1.67 eV) is always doubly-degenerate.\\   
For the case of a positively charged As vacancy, the picture unfolds in quite a different way. 
The effect of a ${\rm V}_{\rm As}^+$ on a pair of Mn dopants at different separations is plotted in Fig.~\ref{fig:2Mn-p}.
In Fig.~\ref{fig:2Mn-p}(a) we plot the superimposed electronic level structure inside the GaAs 
gap for a ${\rm V}_{\rm As}^+$ and a Mn pair
calculated separately, and use it as before as a reference, 
effectively representing the two defects placed at very large
separation. 
Figs.~\ref{fig:2Mn-p}(b)-(f) show the evolution of the electronic 
structure for the coupled Mn-pair--${\rm V}_{\rm As}^+$ system 
when the two defects approach each other.
We find that the two doubly-degenerate levels of the ${\rm V}_{\rm As}^+$ move progressively toward the conduction band under the effect 
of the Mn-pair potential. In contrast, five of the six levels of the Mn pair, including the lowest of the two Mn-acceptor levels 
(the lowest unoccupied 
level in Fig.~\ref{fig:2Mn-p}(a)) are pushed down towards the valence band. The mechanism responsible for these two combined effects 
is essentially the same as for the case of one individual Mn-impurity in the presence of a ${\rm V}_{\rm As}^+$. 
The repulsive potential coming from the Mn ion
pushes away electrons and tends to localize holes closer to the Mn core, 
increasing the energy of the As levels.  Similarly the attractive potential (for the electrons)
coming from the ${\rm V}_{\rm As}^+$ tend to delocalize the Mn-acceptor, 
hereby pushing their energies toward the valance band.
Interestingly, the top Mn-acceptor (the highest unoccupied Mn level in all panels of Fig.~\ref{fig:2Mn-p}) 
is the only one that defies this rule: way up in energy and close to the conduction band edge,
this level is hardly affected by the presence of ${\rm V}_{\rm As}^+$, even at the shortest distances. 
This has important implications on the relative splitting of the two Mn-acceptor levels, discussed below.\\
\begin{figure}[htp]
\centering
\includegraphics[scale=0.6]{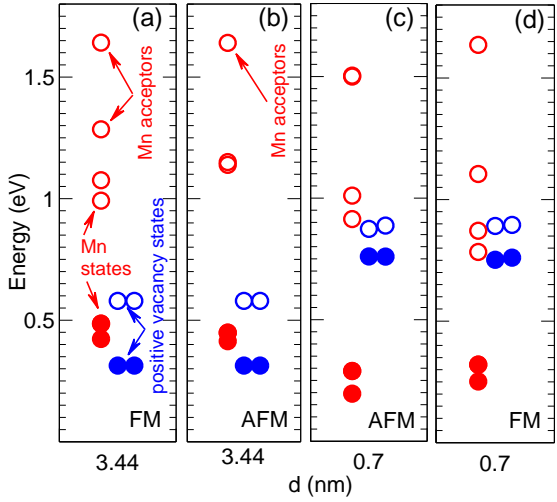}
\caption{(Color online)  The level structure of a pair of Mn impurities and 
a positively charged As vacancy on the GaAs (110) surface for two different defect separations. 
See caption of Fig.~\ref{fig:1Mn-n}.
(a) and (d): The spins of the two Mn atoms are ferromagnetically (FM) coupled.
(b) and (c): The spins of the two Mn atoms are antiferromagnetically (AFM) coupled.
Blue color indicates vacancy states and red color Mn states. 
Filled and empty circles are occupied and empty states, respectively.
The topmost Mn-acceptor level for the AFM case is doubly-degenerate.}
\label{fig:2Mn-p-FM-AFM}
\end{figure}
The energy level occupancy for the energy levels of the coupled system remains the same at all separations. 
The main difference with
respect to the reference state (Figs.~\ref{fig:2Mn-p}(a)) is the double occupancy of the lowest ${\rm V}_{\rm As}^+$ level (which is empty for an
isolated ${\rm V}_{\rm As}^+$) at the expenses of the two Mn levels (occupied for the isolated pair) located around 1 eV.
As mentioned before, it is likely that these results are strongly altered by on-site electronic correlations which suppress
the double occupancy probability.
In any case, our numerical calculations  show that the two Mn-acceptors (i.e., the two topmost levels of the Mn pair) 
are always unoccupied, as for the case of an isolated Mn pair,
regardless of the position of the As vacancy.
This result, involving levels high up in energy and separated from the other impurity levels, 
should be rather robust against on-site correlations.
Furthermore, as Fig.~\ref{fig:2Mn-p-FM-AFM} shows, the energies of these two levels are split when the spins of the 
two Mn atoms are aligned ferromagnetically, but essentially degenerate for antiferromagnetic coupling. 
Therefore, we reach the important conclusion that the splitting of the two Mn-acceptor energy levels, $\epsilon_{{\rm acc}_1}$ and $\epsilon_{{\rm acc}_2}$,
\begin{align}
\Delta \epsilon_i=\epsilon_{{\rm acc}_1}-\epsilon_{{\rm acc}_2}\;,
\label{spl-ch}
\end{align}
is an indicator of the ferromagnetic state of the Mn pair even in the
presence of ${\rm V}_{\rm As}^+$, as it is 
in the presence of ${\rm V}_{\rm As}^0$ as well as for the isolated pair.
\begin{figure}[htp]
\centering
\includegraphics[scale=0.54]{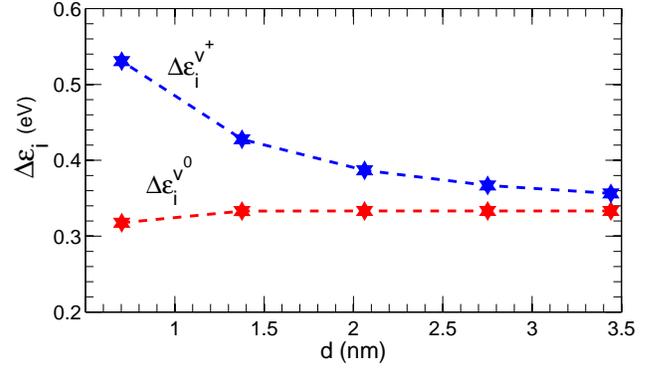}
\caption{(Color online) Change in the energy splitting between the two acceptors levels of a Mn pair (Eq.~\ref{spl-ch}) 
caused by a neutral and a positively charged vacancy placed at a distance $d$ from the Mn pair.
Red and blue asterisks represent the case of a
neutral and a charged As vacancy, respectively.}
\label{fig:splitting}
\end{figure}
In Fig.~\ref{fig:splitting} we plot $\Delta \epsilon_i(d)$ as a function of the Mn-pair--${\rm V}_{\rm As}$ separation, $d$, 
for the cases of a neutral (red asterisks) and a charged As vacancy (blue asterisks). 
We can see that ${\rm V}_{\rm As}^0$ has no 
effect on $\Delta \epsilon_i(d)$, except for 
the shortest separation $d= 0.7$ nm, where
the splitting decreases by merely $\approx 50$ meV. On the other hand our 
calculations show that ${\rm V}_{\rm As}^+$ significantly modifies $\Delta \epsilon_i(d)$,
as ${\rm V}_{\rm As}^+$ is brought closer to the Mn pair and 
$\Delta \epsilon_i(d)$, which is equal to 0.36~eV without any vacancy nearby, 
increases to 0.53~eV at a separation of 0.7~nm.
This figure summarizes our findings on how 
an As vacancy affects the electronic and magnetic properties of a Mn pair in GaAs.\\
Since $\Delta \epsilon_i$ is an indication of the strength of the exchange coupling  between two Mn ions,
these theoretical predictions open up the possibility of manipulating the exchange splitting
by the electrostatic field of nearby As vacancies.\\
\begin{figure}[htp]
\begin{minipage}[h]{1.0\linewidth}
\centering
\includegraphics[scale=0.27]{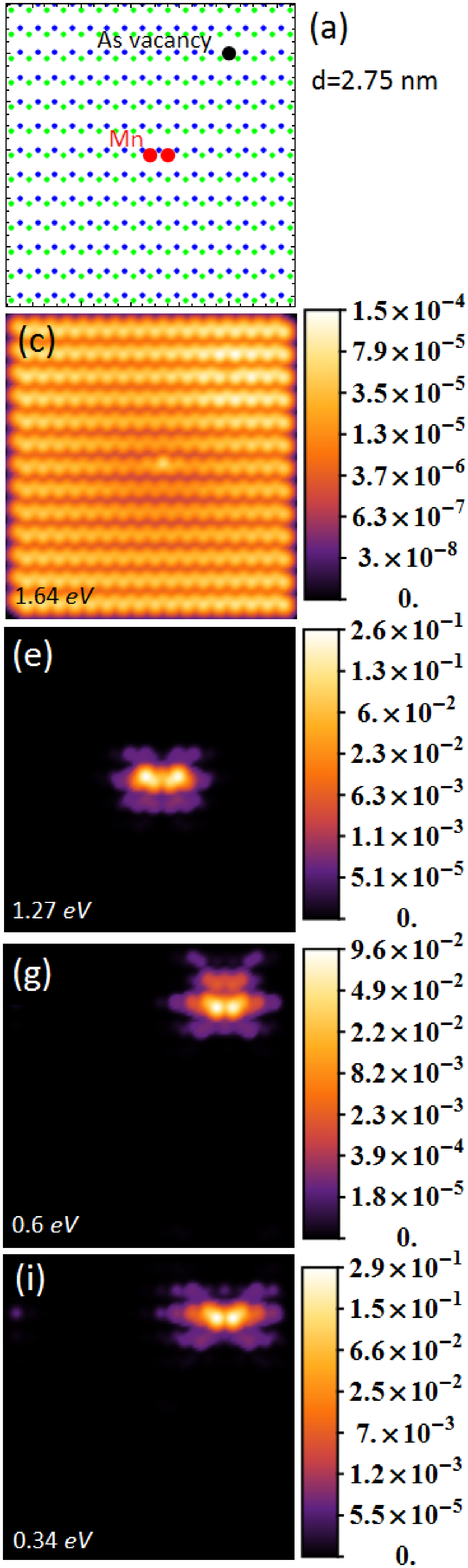}
\hspace{1.50mm}
\includegraphics[scale=0.27]{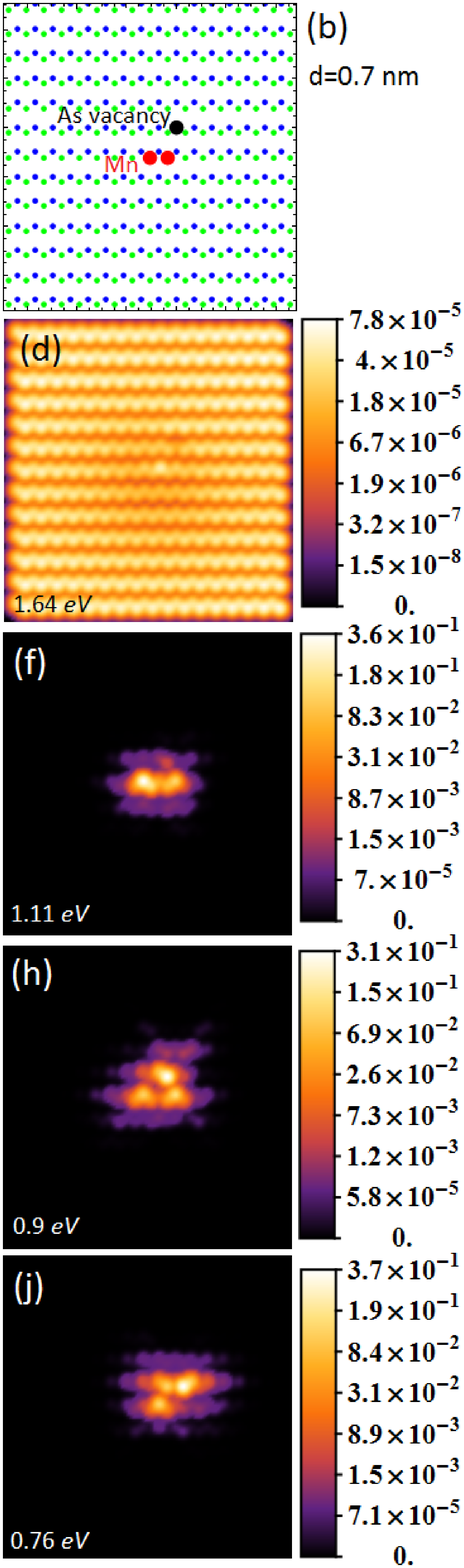}
\end{minipage}
		
\vspace{0.00mm}

\caption{(Color online) 
LDOS of the defect states inside the GaAs energy gap for a system composed of a 
ferromagnetically coupled Mn pair in the presence of a 
positively charged As vacancy on the GaAs (110) surface. Two defect separations are considered:
left column: $d= 2.75$ nm; right column: $d=0.7$ nm. The energy level structure inside the gap for these two
defect configurations is shown in Fig.~\ref{fig:2Mn-p}(c) and  Fig.~\ref{fig:2Mn-p}(f), respectively.
(a) and (b)  Position of the two defects on the GaAs (110) surface for left and right columns, respectively.
(c) and (d)  LDOS of Mn-pair acceptors at 1.64~eV.\
(e) and (f)  LDOS of Mn-pair acceptors at 1.27 and 1.11~eV.\ 
(g) and (i)  LDOS of the two ${\rm V}_{\rm As}^+$ energy levels 
in Fig~\ref{fig:2Mn-p}(c). Each level is doubly-degenerate and two corresponding states have similar LDOS.
(h) and (j): LDOS of the two ${\rm V}_{\rm As}^+$ energy levels in Fig~\ref{fig:2Mn-p}(f).
Each level is a quasi doubly-degenerate level (they are both slightly split by a few meV); 
(h) and (j) represent the combined LDOS of the two corresponding states.
The energy resolved LDOSs for these two pairs of states are plotted in Fig~\ref{fig:2Mn0.7p-spl}.}
\label{fig:2Mn-p-L}
\end{figure}
\begin{figure}[htp]
\includegraphics[scale=0.3]{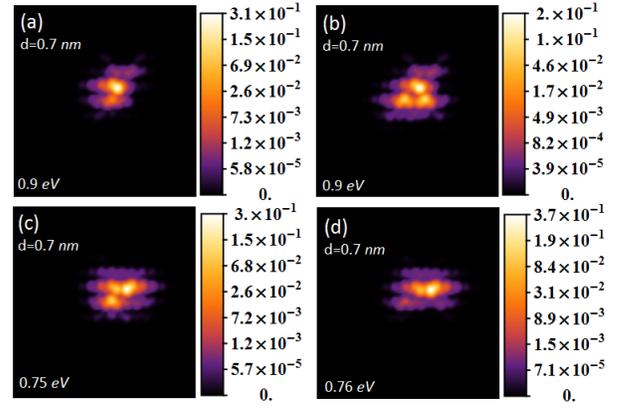}
\caption{(Color online) Energy-resolved LDOS  for the two pairs of states 
corresponding to the two ${\rm V}_{\rm As}^+$ quasi-degenerate 
energy levels at 0.9~eV and 0.76~eV in Fig~\ref{fig:2Mn-p}(f).
Both levels are in fact slightly split, by 4~meV and 9~meV, respectively.}
\label{fig:2Mn0.7p-spl}
\end{figure}
In the last part of this section, we discuss the calculated LDOS for a Mn pair in 
the presence of an As vacancy, which can provide useful
information for ongoing STM experiments. As an example, we consider the case of ${\rm V}_{\rm As}^+$. 
In Fig.~\ref{fig:2Mn-p-L}  we display the LDOS for the defects states inside the GaAs
gap for two defect separations, $d=2.75$ nm (left column) and $d=0.7$ nm (right column).
The electronic energy levels for these two cases were given in  Fig.~\ref{fig:2Mn-p}(c) and  Fig.~\ref{fig:2Mn-p}(f), respectively.
In Figs.~\ref{fig:2Mn-p-L}(a) and (b) we show the positions of the Mn-pair (red circles) and ${\rm V}_{\rm As}^+$ (black circle)
on the GaAs (110) surface for
the two defect separations. 
Figs.~\ref{fig:2Mn-p-L}(c)-(d) and (e)-(f) show the LDOS of the highest and lowest acceptors, respectively.\\
The LDOS of the highest acceptor shows
a quite delocalized state, since, as mentioned above, for the present choice of the TB parameters, this level,
located at 1.67 eV, has merged with the conduction band. 
By tuning the value of $V_{\rm Corr}$ in Eq.~\ref{hamiltonian}, it is possible to bring this level back inside the top of the band gap,
in the position where it is seen experimentally ($\approx 1.03$~eV). 
In this case, the corresponding wavefunction not only is strongly
localized around the Mn, but it also has a bonding character, 
with maximum spectral weight on the As atom between the Mn atoms\cite{yazdani_nat06,scm_MnGaAs_paper2_prb2010}. 
The bonding character for the highest acceptor can be seen, albeit very slightly, in Figs.~\ref{fig:2Mn-p-L}(c)-(d);
although the wavefunction is delocalized, it has a clear maximum on the As between the Mn atoms of the pair.\\
For the lowest acceptor state, the picture is different.
As shown in Fig.~\ref{fig:2Mn-p-L}(e), when the As vacancy is far away, the acceptor state has an anti-bonding
character\cite{scm_MnGaAs_paper2_prb2010}, with $\approx 52\%$ of its total spectral weight
equally distributed on each of the two Mn atoms, and a small contribution on the As in between\cite{scm_MnGaAs_paper2_prb2010}. 
When the As vacancy is at 0.7 nm from the Mn-pair,
Fig.~\ref{fig:2Mn-p-L}(f), 
only $\approx16\%$ of the acceptor spectral weight is located on the Mn atom closest to the As vacancy,
and $\approx36\%$ on the furthest Mn. The rest of the acceptor wavefunction is pushed further away from the vacancy. 
This confirms the repulsion between the ${\rm V}_{\rm As}^+$ and the hole bound to Mn.
Clearly the Mn-acceptor states hybridize very little with the ${\rm V}_{\rm As}^+$ states, even at the shortest
separation, which is consistent with the level structure shown in Fig.~\ref{fig:2Mn-p}(f).
The ${\rm V}_{\rm As}^+$ states, however, behave differently. As shown in Fig.~\ref{fig:2Mn-p-L}(h) and (j) the combined LDOS of the two
${\rm V}_{\rm As}^+$ levels is strongly modified when the defect separation is only $d=0.7$ nm.
A careful analysis of this case shows that the two ${\rm V}_{\rm As}^+$ strongly 
hybridize with other Mn-pair states (not the acceptor!)
that are close in energy, according to Fig.~\ref{fig:2Mn-p}(f). Because of this coupling with the states of the Mn-pair, 
the two ${\rm V}_{\rm As}^+$ levels,
which are essentially doubly-degenerate at  $d=2.75$ nm, are slightly split at the shortest separation (4 meV for the level
at 0.9 eV and 9 meV for the level at 0.75 eV). 
The energy-resolved LDOS of two corresponding states (originating from the same level) now look
very different, as shown in Fig.~\ref{fig:2Mn0.7p-spl}, as a result of the different hybridization with the Mn states.
This is the precisely same phenomenon that we encountered above when studying As vacancies in the vicinity of 
{\it individual} Mn impurities.\\
It is important to point out that the modification of the ${\rm V}_{\rm As}^+$ states due to the hybridization with the Mn states
occurring at very short separations has nothing to do with the shift of the Mn-acceptor levels 
caused by the Coulomb potential of the charged
As vacancy. Indeed, similar hybridization-induced deformations of the LDOS occur also 
in the presence of neutral As vacancies, where the Mn-acceptor energies
are left essentially unchanged.
Nevertheless, the change in the ${\rm V}_{\rm As}$ state wavefunctions caused by nearby Mn impurities (individual or pairs) 
should be detectable in STM experiments.\\
%
%
%
%%%%%%%%%%%%%%%%%%%%%%%%%%%%%%%%%%%%%%%%%%%%%%%%%%%%%%%%%%%%%%%%%%%%%%%%%%%%%%%%%%%%%%
%
%
%HERE
\section{Conclusions}
\label{conc.}
Motivated by STM experiments~\cite{gupta_science_2010}, we have studied theoretically the effect of 
neutral and positively charged As vacancies on the 
electronic and magnetic properties of the acceptor state of 
Mn impurities on the GaAs (110) surface. For the calculations we used an atomistic tight-binding (TB) model in the mean-field approximation
for large finite clusters containing up to 8840 atoms, allowing us to address (110) surface areas on the order of 
$6.6\times 7.1\; {\rm nm}^2$ and
Mn-As separations more than 4 nm.
The model has been quite successful recently in describing the physics 
of individual and pairs of
substitutional Mn atoms on the (110) surface~\cite{scm_MnGaAs_paper1_prb09,scm_MnGaAs_paper2_prb2010}.
For individual isolated As vacancies, both neutral and positively charged,
the model is able to reproduce the electronic structure, explained earlier using {\it ab-initio} 
methods~\cite{chao_prb96, chelikowsky_prl96, Cheli_surf.sci_98},
characterized by two in-gap doubly-degenerate levels, which are both empty for ${\rm V}_{\rm As}^+$, while 
half-filled and empty in the case of  ${\rm V}_{\rm As}^0$. 
These results are remarkable, since they are
obtained within an independent electron approximation
where both types of As vacancies are modeled by effective single-particle potentials,
without requiring self-consistency in the charge distribution.\\
The focus of the paper has been the analysis of the Mn-acceptor binding energy, as a function 
of the separation between the Mn-impurity and ${\rm V}_{\rm As}$. 
The model correctly reproduces the experimental findings that the Mn-acceptor binding energy is unaffected by the presence
of nearby neutral vacancies,  and that the binding energy decreases significantly when a charged As is brought closer to the Mn.
The numerical results supports the hypothesis put forward in 
Ref.~\cite{gupta_science_2010} that the shift in the resonant spectra of Mn 
in STM experiments is not a trivial As-vacancy-induced band-bending effect, but rather a genuine modification of the Mn-acceptor 
level due to the direct Coulomb repulsion of the positively charged Mn-acceptor hole by ${\rm V}_{\rm As}^+$.
Indeed, our calculations of the acceptor local density of states (LDOS)
show that a decrease of the Mn-acceptor binding energy by a nearby ${\rm V}_{\rm As}^+$
is accompanied by increased delocalization of the acceptor-hole wavefunction.
We stress that
in order to investigate Mn-impurity--${\rm V}_{\rm As}$  separations close to the ones studied experimentally ($\approx 4$~nm), large
supercells are needed. This gives our TB model an advantage over {\it ab-initio} methods based on DFT, which typically have 
much smaller maximum supercell sizes. \\
The calculations also show that the electronic energy level structure of the coupled Mn-impurity--${\rm V}_{\rm As}$ system is
considerably more complex than the electronic structure of isolated individual defects and 
less suitable for independent-electron approximation treatments. This limitation applies equally well to our TB 
approach and to approaches which involve solutions of the Kohn-Sham equations of density functional theory.
In particular,
our model predicts the possibility of electron charge switching between 
the levels of the two defects as a function of their separation,
leading to a modification of the in-gap resonant spectra in STM experiments, 
and the appearance of new transport channels. 
We emphasize that these results are obtained within 
a non-self-consistent one-particle theory, which neglects Coulomb correlations.
Such correlations are important in determining the mean level occupancy, and likely to lead to 
Correlation-induced fluctuating valences on both defects.
The coupling of the Mn-impurity and ${\rm V}_{\rm As}$ states
inside the gap gives rise to strongly hybridized wavefunctions at short separations, 
and the LDOS at the position of each 
defect is deformed by the presence of the other. These effects should be visible in STM experiments when
the defect separation is smaller than 1 nm.\\
In the second part of the paper we have investigated the effect of ${\rm V}_{\rm As}$ on pairs of ferromagnetically coupled
Mn atoms. Here our calculations predict that the energies of the two unoccupied Mn-acceptor levels, which are split if the spins of 
the two Mn atoms are parallel, move further apart when a ${\rm V}_{\rm As}^+$ approaches the Mn-pair.
We have shown that the mechanism behind this effect is essentially related to the reduction of the Mn-acceptor binding energy
present in the case of individual Mn impurities.
Consistently with this result, a neutral ${\rm V}_{\rm As}$ does not have any effect on the acceptor splitting.
Previous experimental~\cite{yazdani_nat06} and theoretical work~\cite{yazdani_nat06, tang_spie2009, scm_MnGaAs_paper2_prb2010} 
has shown that the acceptor splitting of the Mn-pair, which appears as a double in-gap resonance
in STM measurements, is an indication of the strength of the exchange interaction between the two magnetic atoms.
The results presented here suggest that it might be possible to control and manipulate the Mn exchange coupling in GaAs 
with local electric fields
at the atomic scale, by means of As vacancies positioned nearby with atomic precision. Since the charge state of the As vacancy can be
reversibly swapped between neutral and charged by the STM tip, this effect could be used to rapidly turn on and off locally
the exchange coupling between the two magnetic atoms.\\ 
Note that in this paper  we have only considered the case of two Mn atoms placed at the shortest
separation from each other. This choice maximizes the acceptor splitting and the exchange interaction.
It is known that the exchange interaction on the (110) surface is strongly anisotropic 
with respect to the orientation along the crystal axis\cite{yazdani_nat06, tang_spie2009, scm_MnGaAs_paper2_prb2010}.
A careful choice of both the Mn-pair orientation and the As vacancy location could result in situations 
where the local exchange coupling can be 
rapidly switched
in real time between two quite different values.
\begin{acknowledgments}
This work was supported by the Faculty of Technology
at Linnaeus University, by the
Swedish Research Council under Grant Number: 621-2010-3761, 
by the NordForsk research network 080134 ``Nanospintronics: theory and
simulations", by the Welch Foundation under Grant No. TBF1473,
and by the DOE Division of Materials Sciences and Engineering under grant No. DE- FG03-02ER45958. Computational resources have
been provided by the Center for Scientific and Technical Computing LUNARC at
Lund University. We are very grateful to
J. A. Gupta for useful discussions and illuminating explanations of the experimental results.
\end{acknowledgments}

%\nocite{*}

\bibliography{As_vac}% Produces the bibliography via BibTeX.
%\bibliography{refs}% Produces the bibliography via BibTeX.

\end{document}